\documentclass[12pt]{iopart}
\usepackage{iopams}
\usepackage[sort&compress,numbers]{natbib}  %for iopart-num
\eqnobysec

\usepackage[utf8]{inputenc}
\usepackage{amssymb}
\expandafter\let\csname equation*\endcsname\relax
\expandafter\let\csname endequation*\endcsname\relax
\usepackage{amsmath}
\usepackage{graphicx}
\usepackage{color}

\newcommand{\be}{\begin{equation}}
\newcommand{\ee}{\end{equation}}
\newcommand{\bea}{\begin{eqnarray}}
\newcommand{\eea}{\end{eqnarray}}
\newcommand{\bean}{\begin{eqnarray*}}
\newcommand{\eean}{\end{eqnarray*}}

\newcommand{\bk}{{\mathbf k}}

\newcommand{\bp}{{\mathbf p}}

\newcommand{\bx}{{\mathbf x}}

\newcommand{\bn}{{\mathbf n}}
\newcommand{\bv}{{\mathbf v}}

\newcommand{\PPP}{{\cal P}}
\newcommand{\RR}{{\cal R}}
\newcommand{\al}{\alpha}

\newcommand{\de}{\delta}
\newcommand{\De}{\Delta}
\newcommand{\ep}{\epsilon}
\newcommand{\ga}{\gamma}

\newcommand{\La}{\Lambda}
\newcommand{\la}{\lambda}
\newcommand{\Om}{\Omega}
\newcommand{\om}{\omega}
\newcommand{\si}{\sigma}

\newcommand{\vth}{\vartheta}
\newcommand{\vph}{\varphi}
\newcommand{\ra}{\rightarrow}

\newcommand{\bm}[1]{\mbox{\boldmath $#1$}}

\newcommand{\dd}{\partial}

\def\id{{\rm 1\kern -2.5pt I}}

\begin{document}
% about 40 pages!

\review[The Cosmic Microwave Background]{The Cosmic Microwave Background:\\
The history of its experimental investigation and its significance for cosmology} 
\author{Ruth Durrer}
\address{Universit\'e de Gen\`eve, D\'epartement de Physique Th\'eorique, 1211 Gen\`eve, Switzerland}
\ead{ruth.durrer@unige.ch}
\date{\today}
%\notitlepage

\begin{abstract}
This review describes the discovery of the cosmic microwave background radiation in 1965 and its impact on cosmology in the  50 years that followed. This discovery has established the Big Bang model of the Universe and the analysis of its fluctuations has confirmed the idea of inflation and led to the present era of precision cosmology. I discuss the evolution of cosmological perturbations and their imprint on the CMB as temperature fluctuations and polarization.  I also show how a phase of inflationary expansion generates fluctuations in the spacetime curvature and primordial gravitational waves.
In addition I present  findings of CMB experiments, from the earliest to the most recent ones. The accuracy of these experiments has helped us to estimate the parameters of the cosmological model with unprecedented precision so that in the future we shall be able to test not only cosmological models but  General Relativity itself on cosmological scales.
\end{abstract}
%\pacs{04.70.Bw, 04.20.-q, 04.70.-s, 97.60.Lf}

\submitto{\CQG}
\maketitle

\section{Historical Introduction}\label{s:intro}
The discovery of the Cosmic Microwave Background (CMB) by Penzias and Wilson, reported in Refs.~\cite{Penzias:1965wn,Dicke:1965zz}, has been a 'game changer' in cosmology. Before this discovery, despite the observation of the expansion of the Universe, see~\cite{mcmc:2015}, the steady state model of cosmology still had a respectable group of followers. 
However, if the 'excess antenna temperature' measured by Penzias and Wilson isotropically in all directions~\cite{Penzias:1965wn} was correctly interpreted by the preceding paper in the same issue of the Astrophysical Journal~\cite{Dicke:1965zz}, the Universe was clearly adiabatically expanding and cooling as postulated by 
Lema\^\i tre~\cite{Lemaitre:1927zz}\footnote{This original reference is in French. One can also read~\cite{Lemaitre:1931zza}, the English translation by Eddington, which, however, omits the important estimate of the Hubble constant and the discussion of the age problem.}  using a solution of Einstein's field equation found previously by Friedman~\cite{Friedman:1922kd}. In 1978, Penzias and Wilson were rewarded with the Physics Nobel Prize for their discovery. 

In the Big~Bang model, the Universe starts out from a hot, dense initial state and subsequently expands and cools. It had been noted already some time ago by Gamow and collaborators~\cite{Gamov:1948,Alpher:1948,Alpher:1948b}, that the Big~Bang model predicts a background of cosmic radiation, a relic from the hot early phase. Its temperature had been estimated to be of the order of  a few degrees Kelvin (in the above papers values from 5K to 50K can be found). The discovery by Penzias and Wilson indicated a CMB temperature of 3K.

After the discovery of the CMB, the Big Bang model of cosmology was established. Together with the observation and explanation of the cosmic abundance of light elements, especially $^4$He~\cite{Gamow:1946eb,Alpher:1948ve}, it strongly indicates that the Universe was much hotter and denser in the past.  The Hubble expansion law, see~\cite{mcmc:2015} for details, 
predicts that also the wavelength of photons expands so that they are redshifted. The redshift $z$ denotes the relative difference of the wavelength at the observer, $\la_o$, to the wavelength of the emitter, $\la_e$, i.e.,   $z=(\la_o-\la_e)/\la_e$.
The energy density of the Universe was actually dominated by the contribution from CMB photons at $z\gtrsim 4000$, i.e. $T\gtrsim  10^4K\simeq 0.93$eV.

\begin{figure}[ht]
\begin{center}
\includegraphics[width=9.5cm]{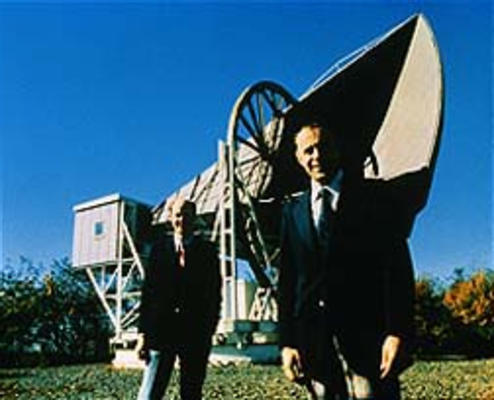}
\end{center}
\caption{\label{f:1}Arno Penzias and Robert Wilson in front of their radio telescope. the most sensitive and modern radio telescope in 1965.}
\end{figure}

The history of Arno Penzias and Robert Wilson is quite amusing (see acount by A. Penzias and by R. Wilson in~\cite{BigBang:2009}). These two young radio astronomers employed by Bell Laboratories at Holmdel, New Jersey, were observing the sky with a radio telescope which had been built to investigate radio transmission from  communication satellites. They had the most advanced radio receiver of the time, a so called horn antenna, see Fig~\ref{f:1}, with a 'cold load' cooled with liquid Helium to suppress interference with the detector heat. But despite this they found a persistent, isotropic receiver noise which was significantly 
larger than what they had expected. 
Also after checking their equipment thoroughly and removing a "white dielectric" (pigeon droppings), this mysterious background noise which corresponded to an antenna temperature of about 3.5K at 7.35 cm would not disappear. 

After a discussion at the phone, a friend, (the radio astronomer B.F. Burke from MIT) sent Penzias a preprint  by Jim Peebles from Princeton University  predicting a cosmic background radiation. Penzias then called Robert Dicke in Princeton and told him that he had measured  'an excess antenna temperature' of about 3K. Dicke, together with Peter Roll and David Wilkinson  visited Bell Labs to see the data and the details of the experiment. When Dicke was convinced they had a result,  Penzias suggested to him that they write a paper together, but Dicke declined~(A.A. Penzias in~\cite{BigBang:2009}).

They finally decided to publish two separate papers back to back. The first by
R.H. Dicke, P.J.E. Peebles, P.G. Roll and D.T. Wilkinson~\cite{Dicke:1965zz} with the title "Cosmic Black-Body Radiation" which interprets the findings as the cosmic background radiation, the CMB, a signature of the hot Big Bang, and the second by A.A. Penzias and R.W. Wilson~\cite{Penzias:1965wn} with the modest title "A Measurement Of Excess Antenna Temperature At 4080 Mc/s".  Here 'Mc/s' are Mega cycles per second hence MHz. This paper reported a "bare-boned account of our measurement -- together with a list of possible sources of interference which had been eliminated" (Arno Penzias in~\cite{BigBang:2009}). For this discovery they were awarded the Nobel Prize in 1978. 

What was the reason for this delay of 13 years? It was certainly not that the cosmological community had not appreciated the importance of their discovery. On the contrary, the Princeton group under R. Dicke had a running experiment at the same time which soon confirmed the discovery by Penzias and Wilson~\cite{PrincetonExp} at 3.2cm and several
experiments showing also the isotropy of the radiation~\cite{PrincetonExp2} followed soon.
 Nevertheless, already Gamow~\cite{Gamov:1948} had predicted that the CMB should be a \emph{thermal}, i.e., a blackbody radiation and this was confirmed once measurements not only in the Rayleigh Jeans part but also in the Wien part of the spectrum, at frequencies $\nu \gtrsim 150$GHz had been made. It took until  the 70s to convince the community that the spectrum was a blackbody and therefore a relict from the Big Bang (see contributions by Robert W. Wilson and R. Bruce Partridge in~\cite{BigBang:2009}).  
% \rut{which paper?}  
 
 Already before the discovery by Penzias and Wilson, excited rotation states of CN (cyanogen) molecules in interstellar space had been observed, first  by McKellar (1940), 
 and then by Adams (1941)~\cite{1941PASP...53..209A,1940PASP...52..187M}, which corresponded to a sky temperature of about 3K.   But they had not been interpreted as due to the CMB.  Now we know that they are excited by CMB radiation and in several publications they have later been used to measure the CMB temperature, see e.g.~\cite{1993ApJ...413L..67R}.

\begin{figure}[ht]
\begin{center}
\includegraphics[width=0.8\linewidth]{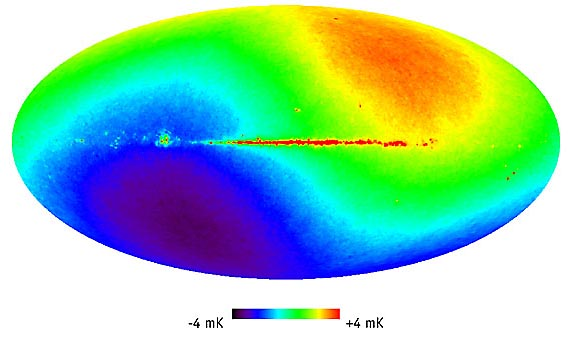}
\end{center}
\caption{\label{f:dipol} The CMB dipole in galactic coordinates as seen by the WMAP satellite. The red horizontal line in the middle is due to emission from the Milky Way. This figure is obtained by subtracting the best fit monopole from the full sky CMB map and by removing the annual modulation shown in Fig.~\ref{f:dipol-modul}. Figure from the WMAP webpage of NASA~{\tt http://map.gsfc.nasa.gov/mission/observatory\underline{~}cal.html}.}
\end{figure}

 As mentioned above, immediately after the discovery by Penzias and Wilson, cosmologists started to look for anisotropies in the CMB radiation. This was motivated by the assumption that structure in the Universe, galaxies, clusters, voids and filaments formed from small initial fluctuations by gravitational instability. If this idea is correct, then these initial fluctuations must also be present in the CMB.  For a long time the searches for anisotropies just revealed a dipole which was first announced in 1969~\cite{Conklin:1969} (see figure~\ref{f:dipol} for a representation of modern dipole data). Only upper limits were reported on smaller angular scales. In the late 80s, when the present author was a graduate student in cosmology, we knew that $(\De T/T) \lesssim 10^{-4}$ and therefore a purely baryonic Universe could not form the observed structure after decoupling from the CMB radiation. Since baryons can only start clustering once they decouple from the photons and becomes pressureless, there is simply not enough time for so small initial fluctuations to grow to form the observed structures. Dark matter, i.e. particles which do not interact with CMB photons so that their fluctuations can start growing earlier, is needed. Since pressureless matter fluctuations only can start growing once they dominate the energy density of the Universe, a sufficient amount of dark matter is needed. Long before,  Fritz Zwicky had postulated the existence of dark matter in galaxy clusters as the only possibility to explain their large virial  velocities~\cite{1937ApJ....86..217Z}. 
Later, in the 1970s, Rubin et al.~\cite{1970ApJ...159..379R,1978ApJ...225L.107R} introduced dark (non-luminous) matter to explain the flat rotation curves of stars and satellites around galaxies. 
 
 In November 1989 the NASA satellite COBE was launched. It not only measured the CMB spectrum with amazing precision but it also found fluctuations in the CMB on the level of $10^{-5}$.
 After this, the dam was broken and many experiments  were performed on Balloons (e.g Boomerang),
 from earth, especially from the south pole (e.g. ACBAR) but also from the Atacama desert (e.g. ACT),
 and from space, the WMAP and Planck satellites. They revealed not only the temperature anisotropies with high precision but also the slight  polarization which is generated on the last scattering surface by the direction dependence of Thomson scattering.
 
 In the remainder of this review, I shall discuss the results from  these experiments and  their relevance for cosmology. I think it is fair to say that the CMB is the most precious dataset for cosmology. This is not only due to the very precise experiments, but also to the fact that the data can be understood by simple linear cosmological perturbation theory with some non-linear terms added which are well under control.
 
 In the next section I report the discovery of the CMB dipole and its significance. In 
 Section~\ref{s:COBE} the findings from the COBE satellite are  discussed which led to the second Nobel prize given for the CMB. To appreciate the importance of these findings I 
give a brief introduction to  cosmological perturbation theory and to the theory of inflation. I shall not derive the results but only describe them and explain their  physical origin. Mathematical derivations can be found in the original literature or in my book on the subject~\cite{2008cmb.book}.  Section~\ref{s:prec} is devoted to the more recent experiments, mainly the NASA satellite WMAP and the ESA satellite Planck. I  also present a brief introduction to the cosmic history and to cosmological parameters. This is needed to understand why these measurements allow us to determine the cosmological parameters, i.e. the handful of 'arbitrary' numbers which govern the evolution of the Universe, with unprecedented precision. CMB polarization and its significance are discussed in Section~\ref{s:pol} and in Section~\ref{s:futur} an outlook on the future of CMB physics is attempted. In Section~7 I conclude.
 
 At this point I also want to make a disclaimer. There are so many CMB experiments, all of them contributing their essential part to the puzzle, that there is simply not enough space to describe all of them. Also, I think this would not lead to a very entertaining article. Therefore my citations of experiments, apart from the really crucial ones, is somewhat accidental and I apologize if your favorite experiment is not mentioned.\vspace{0.2cm}

{\bf Notation:} In this article the speed of light, Planck's constant and Boltzmann's constant are set to unity, $c=\hbar =k_B=1$. This means that time and length have the same units
which is the inverse of the unit of mass, energy or temperature. The Planck mass is defined by $m_P^2=1/\sqrt{G} =\sqrt{\hbar c/G} \simeq1.22\times 10^{19}$GeV.
 
 %%%%%%%%%%%%%%%%%%%%%%%
% immediate impact: acceptance of big bang
% long term impact: precision cosmology
% remaining questions, current research: r, B-polarization, d_A  and 2nd order lensing.
\section{The CMB dipole}\label{s:dip} 
\begin{figure}[ht]
\begin{center}
\includegraphics[width=0.7\linewidth]{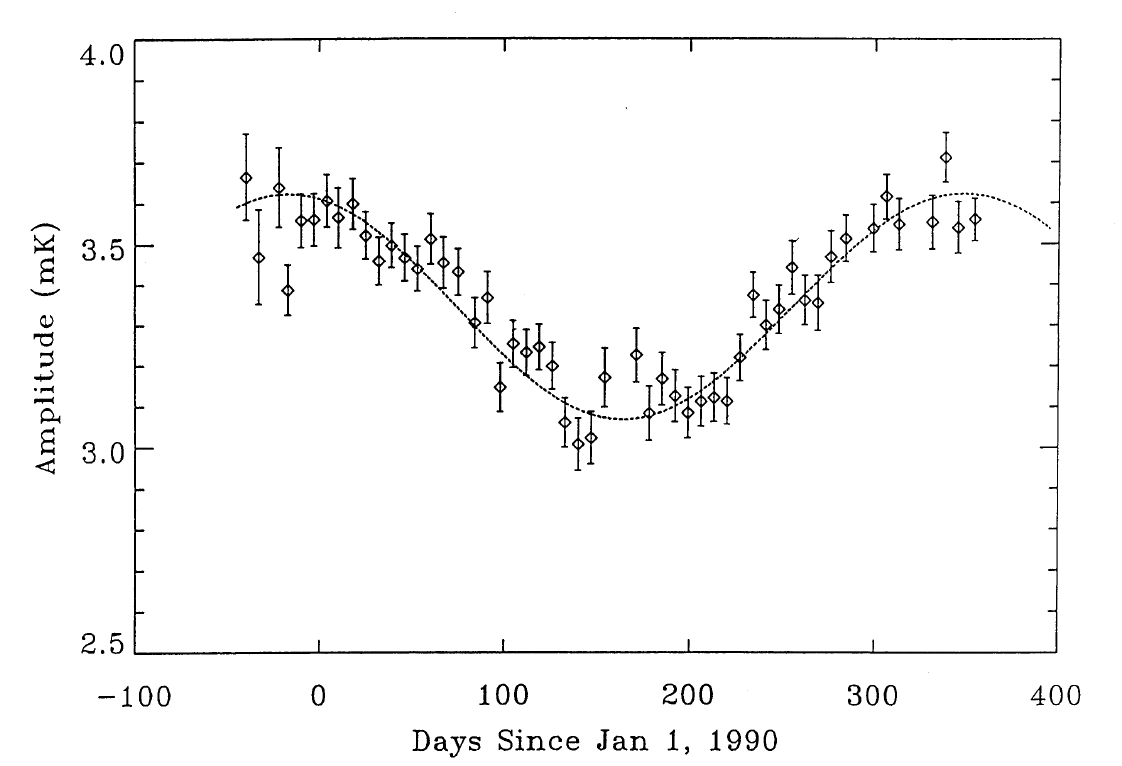}
\end{center}
\caption{\label{f:dipol-modul} The yearly modulation of the CMB dipole with an amplitude of about 0.3mK, due to the motion of the earth. The expected theoretical curve is also indicated. Figure from~\cite{1992ApJ...391..466B}.}
\end{figure}

In 1969 Conklin~\cite{Conklin:1969} and soon after that Henry (1971)~\cite{Henry:1971}, reported the first finding of  a dipole anisotropy in the CMB. Later, Corey and Wilkinson (1976)~\cite{1976BAAS....8Q.351C} performed a more precise and detailed experiment with better error control which also detected the dipole.
The latest measurements of the dipole have been reported by the satellite experiments COBE~\cite{Kogut:1993ag} and WMAP~\cite{Hinshaw:2008kr}.
The latest value is 
\be
\left(\frac{\De T}{T}\right)_{\rm dipole} =~ (1.2312 \pm 0.0029  ) \times 10^{-3} \,.
\ee
Here $T$ is the photon temperature and $\De T$ is its fluctuation amplitude on a scale of $180^o$.
We interpret this value as due to our proper motion with respect to the surface of last scattering. 
Indeed, an observer moving with velocity $\bv$ relative to a source
in direction $\bn$ emitting a photon with proper momentum $\bp=-\ep\bn$ sees
this photon red- (or blue-) shifted with frequency $\nu'=\ep'/h$ where
\be
\ep'=\ga\ep\left(1 + \bn\cdot\bv\right) , \qquad \gamma=\frac{1}{ \sqrt{1-v^2}} \,. 
\ee
(Here and in the rest of this article the speed of light is set to unity, $c=1$.)
For an isotropic emission of photons coming from all directions $\bn$ at first order
in $\bv$ this leads to a dipole anisotropy. 
Interpreting it as due to our motion with respect to the last
scattering surface  implies a velocity for the bary-center of the solar-system
 given by
\be
v=369\pm0.9 {\rm km/s} ~~ \mbox{ in direction } (b,l)=(48.26^o\pm 0.03^o,263.99^o\pm 0.14^o)
\ee
at 68\% CL \cite{Kogut:1993ag,Fixsen:1996nj,Hinshaw:2008kr}. Here $(b,l)$ denote the latitude ('Breite') and longitude ('L\"ange') in galactic coordinates. Interestingly, COBE not only measured the amplitude of the dipole with high precision but also its annual modulation due to the motion of the earth
which moves with a mean velocity of about 30km/s with respect to the bary-center of the solar system~\cite{1992ApJ...391..466B} see Fig.~\ref{f:dipol-modul}.

The FIRAS experiment aboard the COBE satellite also measured the frequency spectrum of the dipole and showed that it is in good agreement with the derivative of a Planck spectrum~\cite{Fixsen:1996nj}.

Furthermore, the Planck satellite~\cite{planck} has measured the aberration which is of order $v^2$ and the modulation  of CMB fluctuations on smaller scales due to our peculiar motion.  Consistent values for the velocity of the solar system could be derived also from these effects~\cite{Aghanim:2013suk}.  

As we shall see in the next section, all higher multipoles of the CMB anisotropies are much smaller than the dipole. This supports the interpretation that the dipole is (mainly) due to our peculiar motion with respect to the surface of last scattering.

The standard cosmological solutions to Einstein's equation are homogeneous and isotropic with respect to a congruence of geodesic observers whose proper time is called  cosmic time. Therefore they single out a reference frame, the one at rest with respect to such a cosmic observer. Hence cosmology spontaneously breaks invariance under boosts while it preserves rotational and translational symmetries.  This is not surprising, as there are only three spacetimes which preserve all ten Lorentz symmetries, Minkowski, de Sitter and anti-de Sitter spacetimes. 

The observed Universe has a preferred frame and our solar system moves with 369km/s with respect to this frame. This motion has also been approximately confirmed by studying the dipole of far away supernovae~\cite{Bonvin:2006} and the galaxy distribution, 
see~\cite{2012MNRAS.427.1994G} for a review of the dipoles measured so far.

In cosmology,  'absolute space' is back and it is given by the reference frame of the CMB.
Even though the theory of General Relativity is of course Lorentz invariant, most of its solutions, among them also the cosmological ones, are not. In cosmology, 'motion' and 'rest' do have an absolute meaning.
However, on small scales, where the curvature of spacetime can be neglected, this breaking of Lorentz invariance is irrelevant and special relativity is confirmed with high accuracy, e.g., in laboratories like CERN. For example at LHC where protons are accelerated to
energies $E \simeq 3{\rm TeV}$, hence $\gamma \simeq 3000$ which implies $v=0.9999999$ and even more at the older LEP where electrons and positrons had been accelerated to $100$GeV yielding $\ga\simeq 2\times 10^5$ and $v\simeq 1-1.3\times 10^{-11}$, special relativity is perfectly valid.  

\section{COBE and signs of inflation}\label{s:COBE}
Before we can discuss and appreciate the findings of the satellite COBE (COsmic Background Explorer) we have to give a brief introduction to cosmological perturbation theory and inflation.

\subsection{Cosmological perturbation theory}\label{ss:pert} 
The metric of a homogeneous and isotropic Universe is described by one function, the scale factor $a(t)$ and one number, $K$, the spatial curvature,
$$ ds^2 = -dt^2 + a^2(t)\ga_{ij}dx^idx^j \,, $$
where $\ga_{ij}$ denotes the metric of a 3-space of constant curvature $K$.
Einstein's equations relate the evolution of the scale factor to the matter content of the Universe,
\bea
H^2 \equiv \left(\frac{\dot a}{a}\right)^2 +\frac{K}{a^2} &=& \frac{8\pi G}{3}\rho + \frac{\La}{3}\\
\frac{\ddot a}{a} +  \left(\frac{\dot a}{a}\right)^2 +\frac{K}{a^2} &=&  -8\pi G P +\La \,.
\eea
Here $\rho$ and $P$ are the energy density and the pressure of the cosmic fluid, $H$ is the Hubble parameter and $\La$ is the cosmological constant, see also~\cite{mcmc:2015}. These equations are called the Friedmann equations~~\cite{Friedman:1922kd}. They describe a  homogeneous and isotropic Universe.The present value of the Hubble parameter, called the Hubble constant is  $H_0=100 h $km/sec/Mpc where $h\simeq 0.71\pm 0.05$ is a fudge factor absorbing our ignorance of the value of this constant\footnote{The value $h=0.71$ is a mean between different values which are found in the literature and which will be given in section~\ref{ss:prec}}. \\ 1Mpc$=1$Mega parsec $\simeq 3.26\times 10^6$light years $\simeq 1.03\times10^{14}$sec $\simeq 3.1\times 10^{24}$cm. We normalise the scale factor so that it is unity today, $a_0 = a(t_0)=1$, where $t_0$ denotes present time.

At least locally, the true Universe is not perfectly homogeneous and isotropic. But the fluctuations of the CMB temperature are small. It therefore is reasonable to calculate them to first order in cosmological perturbation theory. In longitudinal gauge the perturbed metric can be written as
\be
 ds^2 = -(1+2\Psi)dt^2 + a^2(t)\left[(1-2\Phi)\ga_{ij} +2h_{ij}\right]dx^idx^j \,.
\ee
Here $\Psi$ and $\Phi$ are the  Bardeen potentials and $h_{ij}$ is transverse and traceless, $\nabla^ih_{ij}= \ga^{ij}h_{ij}=0$ which describes gravitational waves. We do not discuss vector perturbations of the geometry as they are usually not generated during inflation or decay during the subsequent evolution.
For a perfect fluid or for quasi-Newtonian matter comprised of non-relativistic particles the two Bardeen potentials are equal and they correspond to the Newtonian gravitational potential.

Einstein's equations relate the metric perturbations to perturbations in the energy momentum tensor of matter which are described by density fluctuations, $\rho=\bar\rho(1+\de)$, peculiar velocity $(u^\mu) = a^{-1}(1, \dd^iV)$ and anisotropic stress, $\Pi_{ij}$ which is the traceless part of the stress tensor. The perturbation of the trace of the stress tensor, i.e. of the pressure is given by $\de P = c_s^2\de\rho$ for adiabatic perturbations. Here $c_s$ denotes the adiabatic sound speed. The linear perturbation equations and their detailed discussion can be found in~\cite{2008cmb.book}.  

Here I just want to introduce also the curvature perturbation $\cal R$ which is the quantity which is usually calculated for inflationary models. In terms of the Bardeen potentials it is given by
\be
-{\cal R} = \frac{2}{3(1+w)}\left[\Psi + H^{-1}\dot\Phi\right] +\Phi \,,
\ee
where $w=\bar P/\bar\rho$ is the equation of state parameter of the cosmological background.
$\cal R$ is proportional to the perturbation of the spatial Riemann curvature in comoving gauge, i.e. in a coordinate system with vanishing peculiar velocity.

Since the perturbation equations are linear, we can decompose each variable into eigenfunctions of the spatial Laplacian which evolve independently. In the case $K=0$ these are simply the  Fourier modes\footnote{In general,  $K\neq 0$, the situation is somewhat more complicated but one also finds a complete set of eigenfunctions of the spatial Laplace operator~\cite{Kodama:1985bj}}. For simplicity, and since observations indicate that $|K|$ is very small, we concentrate on this case from now on.

As we shall discuss below, inflation determines the initial conditions for each Fourier mode, ${\cal R}(\bk, t_{\rm in})$. We assume that also during the generation of the fluctuations, i.e. during inflation, there is no preferred position nor a preferred direction in space.  Therefore, the resulting fluctuations are statistically homogeneous and isotropic. Because of statistical homogeneity, different Fourier modes are independent and we define the initial power spectrum ${\cal P}(k)$ by
\be\label{eq:Pspec}
k^3\langle {\cal R}(\bk, t_{\rm in}){\cal R}^*(\bk', t_{\rm in}) \rangle =\de(\bk-\bk')2\pi^2{\cal P}(k) \,.
\ee
Here $\langle\cdots\rangle$ denotes the statistical expectation value over many realisations of the Universe.  In simple inflationary models fluctuations are Gaussian so that the power spectrum contains all the information.  We assume that the mean of the perturbations vanish. In the case of adiabatic perturbations all components of the cosmic fluid are initially in thermal equilibrium and are perturbed in the same way.  The power spectrum of some perturbation variable $X$ at late time is then given by the initial power spectrum multiplied by a deterministic transfer function $\Theta^2_X$ which depends only on the matter content of the Universe.
For example for the density fluctuations we obtain
\bea
k^3\langle \de(\bk,t)\de^*(\bk',t)\rangle &=& \de(\bk-\bk')2\pi^2P_\de(k,t)\\
P_\de(k,t) &=& \Theta^2_\de(k,t){\cal P}(k) \,.
\eea
For the CMB the situation is somewhat different as the temperature $T$ is a function of direction, not just of position. For and observer sitting in $\bx$ at time $t$ we have
$T(\bx,t,\bn)$ where $\bn$ is the direction of observation. We expand the direction dependence in spherical harmonics,
\be
T(\bx,t,\bn) = \bar T(t)\sum_{\ell}\sum_{m=-\ell}^{\ell} a_{\ell m}(\bx,t)Y_{\ell m}(\bn)
\ee 
Because of statistical isotropy the random variables $a_{\ell m}(\bx,t)$ for different $\ell$'s and $m$'s are not correlated and we find
\be\label{eq:Cldef}
\langle a_{\ell m}(\bx,t)a^*_{\ell' m'}(\bx,t)\rangle  = \de_{\ell\ell'}\de_{mm'}C_\ell(t) \,.
\ee
Statistical homogeneity requires that the result does not depend on the position $\bx$. Usually we want to evaluate the CMB power spectrum $C_\ell$ today, $t=t_0$ and we suppress the time dependence.

For a given initial spectrum of curvature fluctuations, solving Einstein's equation for the evolution of the geometry and the Boltzmann equation for the evolution of the photon distribution function to first order, one can derive a transfer function $\Theta_T(k,\ell)$ such that the scalar part of the temperature fluctuation spectrum today is given by
\be
\frac{\ell(\ell+1)C_\ell}{2\pi} = \int \frac{dk}{k}\Theta_T^2(k,\ell)\PPP(k) \,.
\ee  
 
The transfer function $\Theta_T(k,\ell)$ depends on the cosmological parameters in many ways.  To understand this let us note a few basic results from cosmological perturbation theory. All of them are derived e.g. in~\cite{2008cmb.book}.
\begin{itemize}
\item On super Hubble scales, $k<aH(t)$, the curvature perturbation as well as the Bardeen potentials remain constant.
\item On sub-Hubble scales in a radiation dominated Universe, density fluctuations oscillate with constant amplitude while the Bardeen potentials 
decay like $(aH)^2$ and also oscillate in phase with the density fluctuations. The peculiar velocity has also constant amplitude but oscillates out of phase with the density. In a standard inflationary Universe, the density fluctuations behave like $\de_\text{rad} \propto \cos(c_sk\eta)$ while the velocity behaves as $v_\text{rad} \propto \sin(c_sk\eta)$. Here $\eta$ is conformal time related to the cosmic time by $d\eta=dt/a(t)$.
\item In a matter dominated universe, the Bardeen potentials  remain also constant on sub-Hubble scales. Density fluctuations grow like the scale factor, $\de_m\propto a$ and the peculiar velocity grows like $v_m\propto a^{1/2}$. Matter fluctuations do not oscillate since the restoring force provided by the pressure is negligible.
\item In a universe dominated by a cosmological constant $\La$, curvature and  fluctuations decay and matter density fluctuations freeze in. 
\end{itemize}

Once CMB photons have decoupled from baryons, when protons and electrons have recombined into neutral hydrogen, the photons move along geodesics and are affected by the gravitational potential along their path:
when the gravitational potential changes in time, a photon which falls into a potential will then have to climb out of a deeper (or less deep) potential leading to a net redshift (or blueshift) of the photon energy. Integrating the photon geodesic along the line of sight one 
finds~\cite{Durrer:1990mk,2008cmb.book}
\be\label{e:dTint}
\frac{\De T}{T}(\bn) = \left[\frac{1}{4}\de_g^{(\ga)} +\bm{V}^{(b)}\cdot\bn + (\Psi+\Phi)\right](t_{\rm dec}, \bx_{\rm dec}) +\int_{t_{\rm dec}}^{t_0}\dd_t(\Psi+\Phi)(t,\bx(t))dt \,.
\ee
Here $\de_g^{(\ga)}$ is the radiation density fluctuation in the spatially flat gauge, $\bm V^{(b)}$ is the baryon peculiar velocity and $\Psi$ and $\Phi$ are the Bardeen potentials.
On large scales, the first and the third term in the square bracket together combine to the ordinary Sachs-Wolfe effect~\cite{Sachs:1967} given in Eq~(\ref{e:SW}). The integral is the so called integrated Sachs-Wolfe effect. It is relevant when the gravitational potential is not constant. As
$\Psi$ is constant in a matter dominated Universe (within linear perturbation theory) this term is relevant right after decoupling, when radiation is not very subdominant (early integrated Sachs-Wolfe effect) and at late time, when the Universe becomes dark energy dominated  (late integrated Sachs-Wolfe effect); or for photons moving through non-linear structures (Rees-Sciama effect~\cite{Rees1968}).

On intermediate scales the first and second term combine to the acoustic oscillations. The second term is the Doppler term which is out of phase with the density term. Hence calling the acoustic oscillations 'Doppler peaks' is truly a misnomer as the Doppler term actually has its maximum inbetween the acoustic peaks. On small scales fluctuations are damped by free streaming (Silk damping~\cite{1967Natur.215.1155S}) and by the finite thickness of the scattering surface. These effects are not captured by Eq.~(\ref{e:dTint}) which is an 'instant decoupling' approximation. To study Silk damping one has to solve the perturbed Boltzmann equation.

If there is a primordial background of gravitational waves, $h_{ij}(t,\bx)$, e.g., from inflation, this also leads to a temperature fluctuations via an integrated Sachs--Wolfe effect given by
\be\label{e:dTgw}
\frac{\De T}{T}(\bn) =-\int_{t_{\rm dec}}^{t_0}\dd_th_{ij}(t,\bx(t))n^in^jdt \,.
\ee
The gravitational wave amplitude is constant on super-Hubble scales and decays like $a^{-1}$ inside the Hubble horizon, i.e., when $k/(aH(t)>1$. Therefore, the contribution from gravitational waves is relevant mainly on scales which enter the horizon after decoupling.

Much more detail on cosmological perturbation theory, especially also the derivation of the Boltzmann approach, can be found in 
Ref.~\cite{2008cmb.book}.

Together with a scale invariant initial  spectrum from inflation (see below), linear perturbation theory therefore predicts a flat 'Sachs-Wolfe plateau' on large scales, $\ell\lesssim 100$, acoustic oscillations on intermediate scales $100\lesssim \ell\lesssim 700$ and damping on small scales, $700\lesssim\ell$.

\subsection{Cosmic inflation}\label{s:infla}
For a long time it was considered mysterious that the Universe started out expanding at the same speed or, equivalently, with the same very high energy density in all  points of space even if these had not been causally connected.  As we shall see below, a very hot radiation dominated Universe has the problem that any two points at an arbitrary distance from each other are causally disconnected at sufficiently early time. This is called the horizon problem.

Note that contrary to what is sometimes said, the Universe was not necessarily small at very early times. Actually, if the present Universe is infinite, it was already infinite at the time of the so called 'Big Bang'.  Of course, we shall never be able to prove that the size of the Universe is infinite, we only know that it is  larger than the present Hubble scale $cH_0^{-1}= 3000 h^{-1}$Mpc.  

The horizon puzzle is most evident when considering the CMB: Its temperature is (nearly) the same all over the sky even though, in a pure radiation--matter universe, patches in the sky which are further apart than about $1^o$ were not in causal contact when the radiation was emitted. Related questions are: why is the Universe so big, so flat and so old?  

In 1980  A. Guth~\cite{Guth:1980zm}  came up with an answer to these questions:
Before the hot Big Bang there was a phase of very rapid accelerated expansion. Such a phase can be realised, e.g., if the energy content of the Universe is dominated by the potential energy of a slowly rolling scalar field. 
Already before, A. Starobinsky~\cite{Starobinsky:1979ty} had studied accelerated expansion and the generation of gravitational waves from quantum fluctuations on such a background.

Neglecting curvature, the Friedmann equation for a slowly varying canonical scalar field takes the form
\be\label{e:friedinf}
H^2= \frac{8\pi G}{3}\rho = \frac{8\pi G}{3}\left(\frac{1}{2}\dot\phi^2 +V(\phi) \right) \simeq 
 \frac{8\pi G}{3}V(\phi)  \,.
\ee
Here $\dot\phi^2/2$ is the kinetic term and $V$ is the potential of the scalar field.
For $\phi=$ constant we have $8\pi GV(\phi)/3 =H^2 =$ constant and the solution is simply $a(t)\propto \exp(Ht)$, exponential expansion.  The scalar field $\phi$ is called the 'inflaton'.

During such a phase of slow roll inflation, the causal horizon, i.e. the distance a massless particle can travel in the interval $t_{\rm in}$ to $t$, becomes 
\be
l_H(t)  =a(t)\int_{t_{\rm in}}^t \frac{dt'}{a(t')} \simeq \frac{[\exp(H(t-t_{\rm in}))-1]}{H} \,.
\ee
For  $t\gg t_{\rm in}$ this becomes arbitrarily large. The cosmological problems mentioned above can be solved\footnote{Note that even though the horizon, curvature and age problems are addressed, homogeneity is assumed. At least an initial homogeneous and isotropic patch which is roughly of the size of the Hubble parameter during inflation,  $H^{-1}$, is needed and is then stretched to the size of the presently  observable Universe. Hence inflation does not truly solve the problem of the homogeneity and isotropy of the Universe.}, by invoking a  phase of about 50 to 60 e-folds of inflation, $H(t_\text{fin}-t_\text{in})\gtrsim 50$.  One also finds that during such an inflationary phase the spatial curvature of the Universe, $K/a^2$ is significantly reduced so that we do not have to be surprised that the observed Universe is consistent with $K=0$ (the flatness problem).

We compare this with a matter or radiation dominated universe where $a\propto t^\alpha$ with $\al=1/2$ for radiation and $\al=2/3$ for matter and $t>0$. In this case we find
\be
l_H(t)  =  \frac{1}{1-\al}\left(t-t_{\rm in}\left(\frac{t}{t_{\rm in}}\right)^\al\right) <  \frac{1}{1-\al}t  \quad \mbox{ for}\quad \al<1 \,,
\ee
which remains small, for a given time $t$, independent of $t_{\rm in}>0$, as long as $\al<1$. From the second Friedman equation one finds (see, e.g.,~\cite{2008cmb.book}) that $\al$ is given by the equation of state parameter
$w$ relating the pressure $P$ to the energy density $\rho$ by $P=w\rho$.   Assuming $w$=constant and neglecting curvature we obtain
$$\al = \frac{2}{3(1+w)}  \,. $$
Hence for $w>-1/3$ we have $\al<1$ and the horizon problem appears. 
If $\al>1$ or equivalently $w<-1/3$, the above integral diverges for $t_{\rm in} \ra 0$ 
and we have no horizon problem. Exponential expansion corresponds to the case $w\ra-1$ where $\al \ra \infty$ and the expansion becomes exponentially fast. Inflationary models with $1<\al<\infty$ are called power law inflation.

\begin{figure}[ht]
\begin{center}
\includegraphics[width=0.6\linewidth]{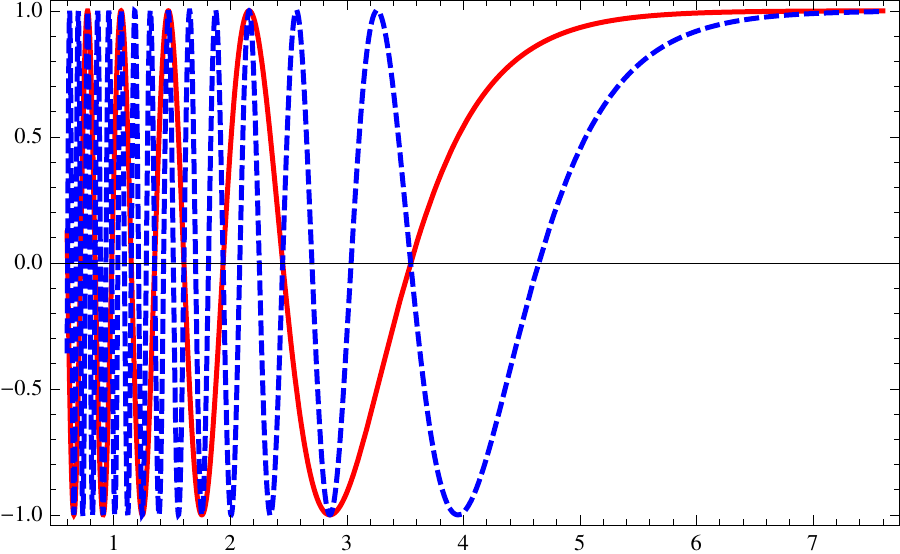}
\end{center}
\caption{The evolution of curvature perturbations during inflation. Before horizon crossing (happening at $t=2$ (red, solid) and at $t=4$ (blue, dashed) in the cases shown), perturbations oscillate as vacuum fluctuations. After horizon crossing they freeze in at constant amplitude.}
\label{f:infla}
\end{figure}

In 1982 Mukhanov and Chibisov~\cite{Muki:1982}  found
that during such a phase of rapid expansion, tiny vacuum fluctuations, which are always present, expand and freeze in. More precisely, they found that the oscillations of vacuum fluctuations of the inflaton field with wavelength 
$\la =a(t)2\pi/k$ are slowed down critically  when it comes into resonance with the expansion rate $k/a(t) \simeq H$. Once $k/a(t) \ll H$, the curvature fluctuations remain constant, see figure~\ref{f:infla}. 

The full calculation, see e.g.~\cite{Slava.book,2008cmb.book}, gives the following result for the power spectrum\footnote{There is a subtlety here that what is calculated are actually vacuum expectation values and which we interpret as statistical expectation values in the late universe. One can show that the disappearance of the decaying mode, 'squeezing' leads to decoherence and hence the fluctuations do become classical soon after inflation~\cite{Polarski:1995jg}.}of the curvature fluctuation, $\RR$, on super Hubble scales
\be
\PPP(k) = \De^2_\RR \left(\frac{k}{k_*}\right)^{n_s-1} \,.
\ee
Here $n_s-1$ is the spectral tilt, $k_*$ is an arbitrary pivot scale and $ \De_\RR$ is the amplitude of the power spectrum at this scale. The spectral tilt and the amplitude depend on the details of the inflationary model.
For the above mentioned models of slow roll inflation one can introduce two slow roll parameters,

\be\label{eq1:slow1}
\ep  \equiv -\frac{\dot H}{H^2}= 
\frac{m_P^2}{16\pi}\left(\frac{V_{,\phi}}{V}\right)^2
\simeq \frac{3}{2}\frac{\dot\phi^2}{V}\ll 1~.
\ee
and
\be\label{eq1:slow2}
\eta \equiv \frac{m_P^2}{8\pi}\left(\frac{V_{,\phi\phi}}{V}\right)  = \frac{V_{,\phi\phi}}{3 H^2}
\,,
\ee
where $V_{,\phi}$ and $V_{,\phi\phi}$ denote the first and second derivative of the potential. Successful inflation requires that both $\ep\ll 1$ and $|\eta|\ll 1$. As soon as one of these parameters approaches unity, slow roll inflation terminates. The calculation of the curvature power spectrum within single field (with canonically normalised kinetic term) slow roll inflation 
gives~\cite{Slava.book,2008cmb.book}
\be
 \De^2_\RR = \left.\frac{H^2}{\pi\ep m_P^2}\right|_{H=k_*/a} \, , \qquad n_s-1 = -6\ep+2\eta \,.
\ee
Here $ |_{H=k_*/a}$ indicates that we have to evaluate the Hubble parameter (which is very slowly varying) when the pivot scale $k_*$ exits the horizon. 

Already before, (in 1979) A. Starobinsky~\cite{Starobinsky:1979ty} had analysed quantum fluctuations of the free gravitational field, i.e., of gravitational waves or tensor perturbations during inflation. These behave very similarly to the inflaton field. They are quantum fluctuations which oscillate at constant amplitude as long as $k/a\gg H$. Once the scale factor has grown sufficiently that $k/a\simeq H$, the oscillations freeze  and when $k/a\ll H$,  $h_{ij}$ becomes  constant.
The power spectrum of  gravitational waves produced in this way during  single field slow roll inflation is given by
\be
k^3\langle h_{ij}(\bk)h^{ij}(\bk')\rangle  = 2\pi^2\delta^3(\bk-\bk') \PPP_h(k) \, , \quad
\PPP_h(k) = \De^2_h \left(\frac{k}{k_*}\right)^{n_t} \,.
\ee
with
\be\label{e:tamp}
 \De^2_h = \left.\frac{16H^2}{\pi m_P^2}\right|_{H=k_*/a} \, , \qquad n_t = -2\ep \,.
\ee
For the tensor to scalar ratio,  $r$,  we obtain
\be\label{e:cons}
r = \frac{\De^2_h}{\De^2_{\RR}} = 16\ep = -8n_t \,.
\ee
This is the so called consistency relation of canonical single field slow roll inflation.

For later we want to stress the following findings:
\begin{itemize} 
\item Inflation predicts a nearly scale invariant spectrum of curvature (scalar) and gravitational wave (tensor) fluctuations. 
\item The amplitude of tensor fluctuations is (nearly) independent of the slow roll parameters; it determines the energy scale of inflation, $E_\text{inf}$. 
Eq.~(\ref{e:tamp}) together with the Friedman equation~(\ref{e:friedinf}) gives
   \be
    \De^2_h = \frac{128V}{3 m_P^4} \simeq \left(\frac{2.6E_\text{inf}}{m_P}\right)^4 \,.
    \ee
\item If tensor fluctuations can be measured and if inflation can be described by a single slowly rolling scalar field, their amplitude and spectral index must satisfy the  slow roll consistency relation~(\ref{e:cons}).
\end{itemize}

The idea is that the inflaton rolls down its potential and finally leaves the slow roll regime and starts oscillating. During these oscillations couplings to standard model particles  lead to the generation of many particles which soon thermalize and the  universe becomes hot with an energy density dominated by relativistic particles, i.e., radiation. This so called reheating process can be rather complicated and it is very model dependent. Apart from the interesting example of Higgs inflation~\cite{Bezrukov:2007ep}, we have no evidence of how the inflaton couples to ordinary matter. It can lead to the formation of topological defects, especially cosmic strings~\cite{Jeannerot:2003qv}, or generate additional gravitational waves on small scales~\cite{GarciaBellido:2007af}. The temperature to which the Universe is reheated depends of course on the energy scale of inflation, but also on the details of the reheating process.

We consider reheating as the true 'hot Big Bang' because during inflation the universe is not in a thermal state and cannot be considered as hot, even if it was in a state of very high nearly constant energy density. Hence reheating should actually be simply called 'heating'. This does not mean that inflation solves the singularity problem. Inflationary models still have singularities in the strict sense of geodesics which cannot be continued to affine parameter $s\ra -\infty$. But, for example in de Sitter space (i.e. for $V=$constant) these singularities are not connected to a diverging energy density like in a standard Friedman Universe. In this sense the singularity can be considered as physically less severe.

After reheating,  the energy density of the Universe is dominated by relativistic particles in thermal equilibrium
and is given by
\be
\rho = \frac{g_*}{2} a_{SB}T^4  \,.
\ee
Here $g_*$ is the number of relativistic degrees of freedom, more precisely $g_* =7N_f/8 + N_b$ ,and $a_{SB}$ is the Stefan-Boltzmann constant. $N_{f,b}$ are number of 
fermionic and bosonic relativistic degrees of freedom respectively. Here a  degree of freedom is called relativistic if the mass of the corresponding particle is smaller than the temperature, $m<T$, which implies that thermal velocities are close to the speed of light.

\subsection{Findings of the COBE DMR experiment}
In 1992 G. Smoot et al.~\cite{Smoot:1992td} published the detection of anisotropies in the CMB on angular scales $\theta\gtrsim 7^o$ which corresponds about to the first 20 harmonics. They found a roughly constant amplitude of $\De T \simeq 1.3\times 10^{-5}T_0$, apart from an anomalously low quadrupole~\cite{Bennett:1994gg}. Even some months before the COBE announcement of April 1992, namely in January 1992, Russian scientists had announced the detection of a CMB quadrupole by the Relikt-1 experiment on board the Prognoz 9 satellite in the range of $6\times 10^{-6}$ to $3.3\times 10^{-5}$ at the frequency of 37GHz~\cite{1992SvAL...18..153S}. Since their value has such large error bars and since Relikt-1 observed only at one frequency, this first detection is not quoted very often. The DMR experiment aboard the COBE satellite measured the CMB at three frequencies, $\nu =31.5$GHz, $53$GHz and $90$GHz with a resolution of $7^o$ and good precision.

The $C_\ell$'s defined in eq.~(\ref{eq:Cldef}) are the CMB power spectrum. They are related to the correlation function by
\be
{\cal C}(\theta)=\langle \De T(\bn)\De T(\bn') \rangle = \frac{T_0^2}{4\pi}\sum_{\ell}(2\ell + 1)C_\ell P_\ell(\mu)\,, \quad \mu=\bn\cdot\bn'=\cos\theta\,.
\ee
Here $\theta$ is the angle between $\bn$ and $\bn'$ and $P_\ell$ is the Legendre polynomial of degree $\ell$.
As one can show, see e.g.~\cite{Sachs:1967,2008cmb.book}, on large angular scales the temperature fluctuations from  adiabatic inflationary initial fluctuations are given by
\be\label{e:SW}
\frac{\De T}{T_0}(\bx_0,t_0,\bn) = \frac{1}{3}\Psi(\bx_{\rm dec},t_{\rm dec}) \,.
\ee
 Here $\Psi$ is the Bardeen potential evaluated at the position $\bx_{\rm dec}$ at which the photon coming in to the observer at $\bx_0$ from direction $\bn$ has left the last scattering surface, and at the time of decoupling, $t_{\rm dec}$,  $\bx_{\rm dec} = \bx_0+\bn(\eta_0-\eta_{\rm dec}) $. This is the ordinary Sachs-Wolfe effect~\cite{Sachs:1967}.

\begin{figure}[ht]
\begin{center}
\includegraphics[width=6cm]{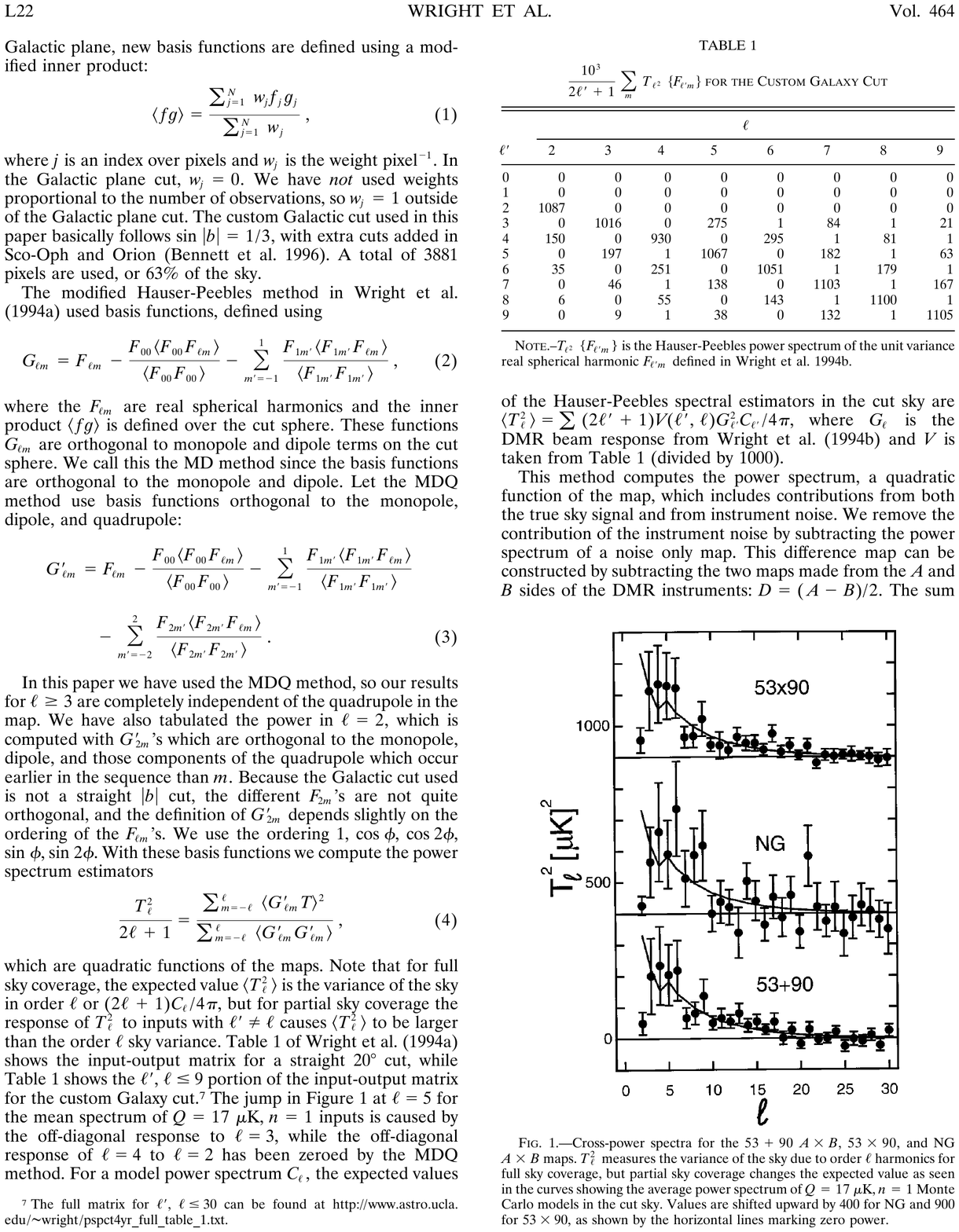}
\end{center}
\caption{\label{f:CoBE-spec}The CMB spectrum from the COBE experiment, $T_\ell^2 =  T_0^2(2\ell+1)C_\ell/4\pi$ using different methods. The spectra denoted by 'NG' and '53x90' are shifted upwards by $400(\mu$K$)^2$ and $900(\mu$K$)^2$  respectively. The solid line is the theoretical result for a scale-invariant, $n_s=1$ power spectrum with quadrupole amplitude $Q=17\mu$K analysed by the same method. Details are found in Ref.~\cite{1996ApJ...464L..21W} from where this figure is reproduced.
}
\end{figure}

In a matter dominated universe, the Newtonian potential is equal to the Bardeen potential which is related to the curvature perturbation $\RR$ by $\Psi = -(3/5)\RR$~\cite{2008cmb.book}. With this the temperature power spectrum on large scales can be related very simply to the curvature power spectrum
\be
\frac{\ell(\ell +1)C_\ell}{2\pi} \simeq \frac{1}{25}\De_\RR^2 \qquad \mbox{for }  n_s\simeq 1\,.
\ee
The DMR experiment aboard COBE measured $\De_\RR^2 \simeq 2\times 10^{-9}$ and $n_s\sim 1$ with still relatively large error bars, e.g., $n_s = 1.21\pm 0.57$. This was the first  confirmation of a prediction from  inflation.
The first published power spectrum reproduced in Fig.~\ref{f:CoBE-spec} is not very clean, but positive power has been found with high significance.
\subsection{The CMB frequency spectrum from COBE}\label{s:spec}
In 2006 the Nobel Prize in physics went jointly to George Smoot, the PI of the DMR experiment and to John Mather, the PI of the FIRAS experiment, both aboard the COBE satellite. The DMR experiment which measured the anisotropies discussed above made measurements at 3 frequencies while the FIRAS experiment had aboard an absolute spectrograph to measure the intensity of the CMB radiation in the interval $30{\rm GHz}
\le\nu\le 600{\rm GHz}$. Especially the data on the Wien part of the spectrum, $\nu> \nu_{\rm peak} \simeq 150$GHz, had been very sparse and imprecise before. The FIRAS experiment measured the CMB temperature with the unprecedented precision 
of~\cite{Mather:1991pc,Mather:1993ij,Fixsen:1996nj,Fixsen:2009ug}
\be
T_0 = 2.72548 \pm 0.00057 K \,.
\ee

\begin{figure}[ht]
\begin{center}
\includegraphics[width=10cm]{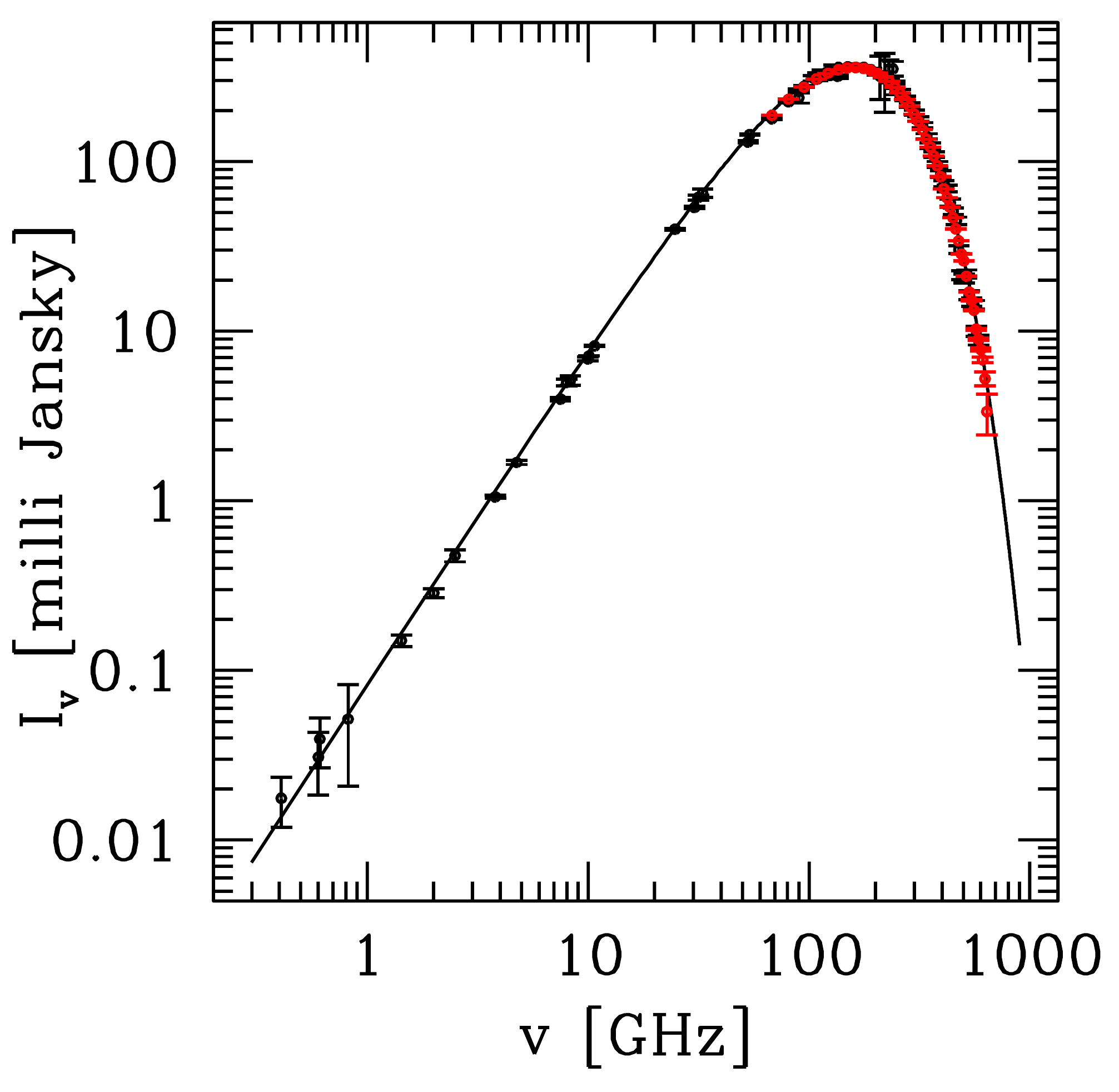} \quad
\caption{The spectrum of the cosmic background
radiation. The data are from many different measurements which are all
compiled in~\cite{Kogut:2006kf}. 
The points around the top (in red) are the measurements from the
FIRAS experiment on COBE~\cite{Fixsen:1996nj}, see also~\cite{Fixsen:2009ug}. The CMB intensity is given in milli Jansky, where $1$Jansky 
$= 10^{-23}$erg/cm$^2$. The line
traces  a blackbody spectrum at a temperature of 2.7255 K (the data is
curtesy of Susan Staggs). Note that for most of the red data points the error bars are smaller than the point size!
\label{f:spec}}
\end{center}
\end{figure}

The FIRAS experiment also put very stringent limits on a chemical potential $\mu$ or a Compton-$y$ distortion which is generated when thermal photons pass through a hot electron gas at a different temperature~\cite{1969Natur.223..721S}. If CMB photons pas through a gas of hot electrons at a temperature $T_e \gg T_{\rm CMB}=T$, the modification of the Planck
spectrum can be determined analytically in terms of a single parameter, $y$, which is given as an integral over the electron density $n_e$ along the line of sight,
\bea
 y &=& \si_T\int n_e\frac{T_e}{m_e}dr \,.\\
 \frac{\de T}{T}(\nu) &=& -y\left[4-\frac{\nu}{T}\coth\left(\frac{\nu}{2T}\right)\right]
   \simeq  \left\{\begin{array}{ll} -2y & \mbox{if  } \nu \ll T \\  y\nu/T & \mbox{if  } \nu \gg T \,. \end{array} \right.
\eea
When passing through a hot plasma, the low energy Rayleigh-Jeans regime of the photon spectrum is depleted and the high energy, Wien part is enhanced. The spectral change vanishes at $\nu_0\simeq 3.8T$ given by $4T=\nu_0\coth(\nu_0/2T)$.

Today, measuring the Compton-y signature in the CMB behind a cluster, the so called Sunyaev-Zel'dovich effect~\cite{1969Ap&SS...4..301Z,1969Natur.223..721S,2008cmb.book}, has become one of the standard methods to detect clusters of galaxies, see~\cite{Ade:2013skr,Bleem:2014lea}. The FIRAS limits on these parameters and on a contribution from free-free emission (Bremsstrahlung) in the CMB radiation are~\cite{Fixsen:1996nj}
\be\label{e:spec-dist}
|\mu| < 9\times 10^{-5}, \quad
|y| < 1.2 \times 10^{-5}, \quad
|Y_{\rm ff}| < 1.9 \times 10^{-5}.
\ee
Here $Y_{\rm ff}$ describes a late time distortion of the CMB given by $(\de T/T)(\nu) = Y_{\rm ff}(T/\nu)^2$ due to free-fee emission from a warm intergalactic medium or from re-ionisation. $Y_{\rm ff}$ can be expressed as an integral over $n_e^2$, see~\cite{Bartlett1991}. Note that these are full sky averages of these parameters, their local values, e.g. the $y$ parameter in the region of a cluster can be significantly larger and have actually been detected as mentioned before.

These are the most stringent limits on distortions of the CMB so far. Since the COBE measurements, no other satellite has measured the CMB spectrum and we have no new information on it since the final analysis of the FIRAS experiment reported 
in~\cite{Fixsen:1996nj}. The data used in this analysis is now 25 years old.

More precise CMB spectral data would be an easy target for a satellite with modern technology and it would be very interesting for several reasons which we shall discuss in Section~\ref{s:futur}. The present information on the CMB spectrum is collected in Fig.~\ref{f:spec}.

\subsection{CMB anisotropies before WMAP}
At the time when the COBE results came out, inflation was not the only mechanism to predict a scale invariant spectrum of CMB fluctuations. Already in the 70ties Harrison and Zel'dovich~\cite{Harrison:1970,Zeldovich:1972} had argued that the only spectrum of fluctuations  that neither leads to black hole formation on small scales nor to large deviation from the observed  homogeneity and isotropy of the Universe on large scales is a scale invariant spectrum.

Furthermore, in 1976 Kibble~\cite{Kibble:1976sj} had proposed that cosmic strings, topological line defects which can form after a symmetry breaking phase transition, might seed the formation of cosmic structure. He showed that their inhomogeneous energy density scales like the energy density of the background universe and therefore always remains the same small fraction of it. It soon became clear that such cosmic strings~\cite{VilShe} and other 'scaling seeds' like global topological defects or self ordering scalar fields also lead to a scale invariant or Harrison-Zel'dovich spectrum of CMB fluctuations,  see~\cite{Durrer:2002} for a review.
It was therefore important to find an observational signature which would distinguish between inflationary fluctuations and topological defects. 

It had been established already in the 70ties~\cite{Peebles:1970ag,1978SvA....22..523D} that the acoustic oscillations of the photon/baryon fluid prior to the decoupling of photons would leave a signature in the CMB anisotropies in the form of so called 'acoustic peaks'. These peaks are very pronounced for inflationary fluctuations but nearly entirely washed out in fluctuations from cosmic defects which are predominantly iso-curvature and for which the phases of a given wavelength are not 
coherent~\cite{Durrer:1995ni,Magueijo:1996px}. Therefore, the detection of the
acoustic peaks was decisive in distinguishing between inflation and topological defects or other scaling seeds.

\begin{figure}[ht]
\begin{center}
\includegraphics[width=11.5cm]{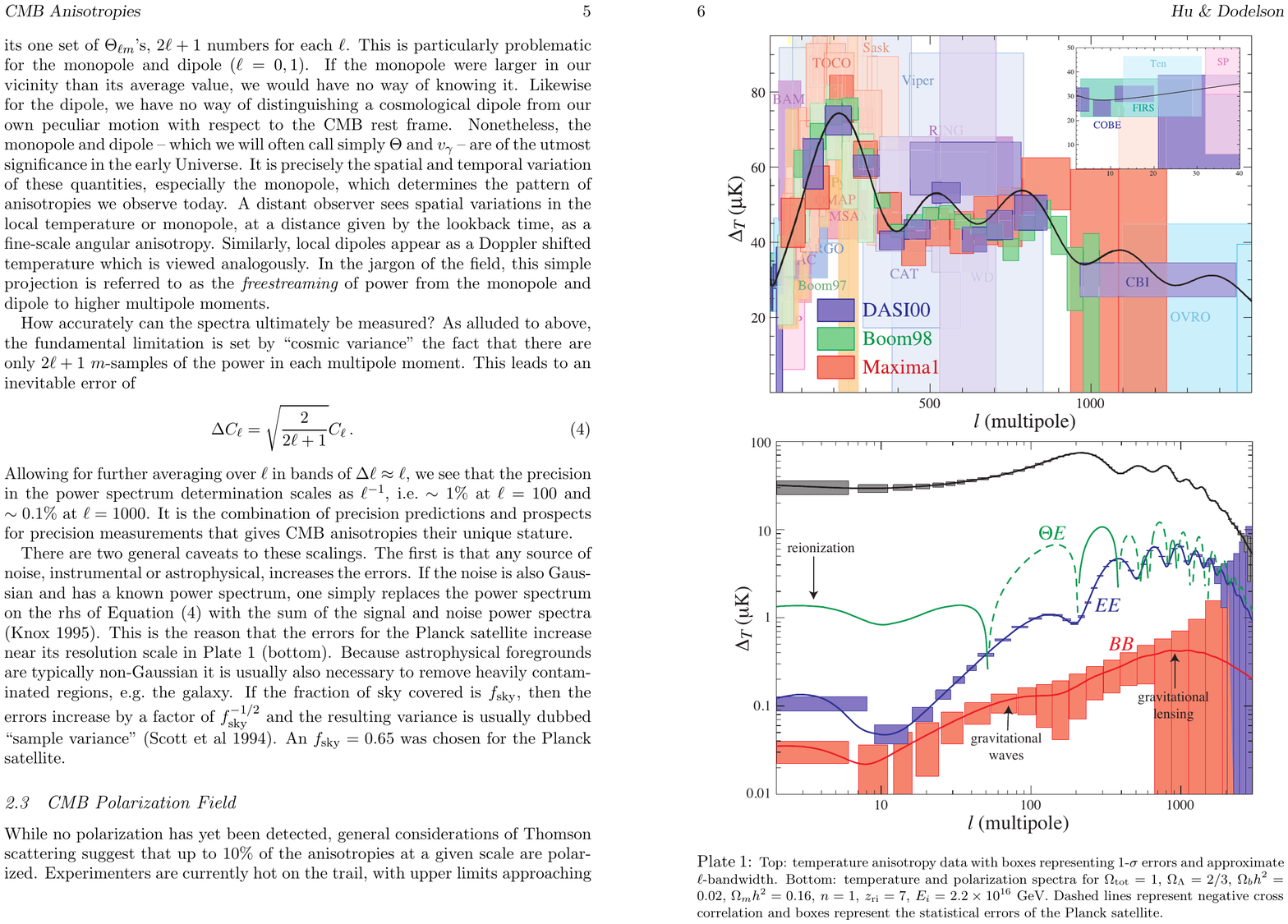}
\end{center}
\caption{\label{f:data2002}The  data on CMB anisotropies before the arrival of the next satellite mission, WMAP (figure from~\cite{Hu:2001bc}).}
\end{figure}

This was achieved especially by the Boomerang~\cite{deBernardis:2000gy}, Maxima~\cite{Stompor:2001xf} and DASI~\cite{Leitch:2001yx} experiments by the end of the last century. A compilation of the situation in 2002, right before
the arrival of the first WMAP results, is shown in Fig.~\ref{f:data2002}. There is clearly a pronounced peak, a signature which is not present in the CMB spectrum of cosmic strings or other topological defects. Even though the data are still 'all over the place' the most precise results confirm the presence of at least one peak.  

\section{Precision cosmology, the WMAP and Planck satellites}\label{s:prec}
WMAP stands for Wilkinson Microwave Anisotropy Probe~\cite{wmap,Hinshaw:2003ex}. The satellite  is named after  David Wilkinson, one of the founding fathers of CMB physics who was also heavily involved in the COBE satellite. He had planned this experiment but passed away at the end of the first year of data-taking after having seen the first results. WMAP is a NASA satellite experiment which was launched in 2001 and took data for 9 years. This was possible since it had only passively cooled elements aboard and hence no need for liquid helium. It took data with radiometers on 5 frequencies from 22.8GHz to 93.5GHz, all of them sensitive also to polarization. The final WMAP results are published in~\cite{Bennett:2013}. The full data is now publicly available.

The ESA satellite 'Planck'~\cite{planck,Ade:2013ktc} was launched in 2009 and took data for about three years.
It was deactivated in October 2013 after three years of nearly flawless operation.
After two years of data taking, the satellite ran out of liquid helium and the HFI (high frequency instrument) ceased functioning. This very sophisticated experiment took data at 9 frequencies from 30GHz to 857GHz, 7 of which are sensitive to polarization. It was composed of bolometers (the HFI) and radiometers (the low frequency instrument, LFI).
It measured the temperature fluctuations with cosmic-variance-limited sensitivity down to an angular scale of a few arc minutes, see~\cite{Ade:2013ktc,Ade:2013kta} for details. Despite its unprecedented sensitivity, unfortunately the instrument was not optimised to measure polarisaton and so we are still waiting for its definite polarization spectrum. On the other side, the unprecedented spectral coverage of the instrument allows for very good foreground rejection. 
\vspace{0.1cm}

Before we can appreciate the meaning of the WMAP and Planck data, we need to briefly discuss the thermal history of the Universe and cosmological parameters, see also~\cite{mcmc:2015}.

\subsection{The thermal history of the Universe}
A photon emitted with wavelength $\la_e$ at time $t$ is received with wavelength $\la_o$ at time $t_0$. The wavelength  expands with the expansion of the Universe such that $z(t)=(\la_o-\la_e)/\la_e =[1-a(t)]/a(t)$, i.e. $z(t)+1=1/a(t)$, remember that we use the normalisation $a(t_0)=1$. Therefore, a cosmic epoch $t$ can also be characterised by its redshift, $z(t)$. Since the photon temperature is inversely proportional to the wavelength we have also $T(t)/T_0 = 1+z$. i.e. at high redshift the Universe is not only denser but also hotter.

\paragraph{Decoupling at $z_{*}\simeq 1090$ : }
 At redshifts  above $z_*$, $z+1 =1/a(t)> 1100$,  $T= T_0\cdot(1+z) > 3000K  \simeq 0.3$eV, there were sufficiently many photons with energies above the hydrogen ionisation energy of 1Rydberg, $\ep_\ga >$ Ry$=13.6$eV around so that protons and electrons could not combine to neutral hydrogen. As soon as a proton and an electron combined, a CMB photon with $\ep_\ga>13.6$eV re-ionized the hydrogen atom. Only once the temperature dropped below 3000K did neutral hydrogen form and the Universe became transparent to CMB photons. The cosmic microwave background is in the literal sense a 'photo' of this time of decoupling. At that time, the Universe was about $t_{*}\simeq 10^{13}(0.14/\Om_mh^2)^{1/2}$sec $\simeq 3\times 10^5$ years old. Note that due to the large entropy of the Universe given by $s\simeq s_\ga \simeq n_\ga \simeq 10^{10}n_b$, this happened at a temperature much below the ionization energy of hydrogen.  
Here $s$ is the entropy density of the Universe and $n_\ga$, $n_b$ are the photon and baryon number densities respectively.
From $T=13.6$eV to $T=0.3$eV, the Universe expanded by a factor of nearly 50 until the photon density in the high energy tail of the Planck distribution with $\ep_\ga>13.6$eV dropped below the baryon density. At a somewhat higher temperature  helium nuclei have already combined with electrons, first to He$+$ and then to neutral helium. 

Much later in the Universe, $z\simeq 10$, hydrogen is reionized by the UV photons from the first stars. This process is poorly understood but is observed in the CMB fluctuation and polarization spectrum as well as in the absence of a Gunn-Peterson trough  in the spectra of quasars with redshift $z<6$. If neutral hdrogen atoms would be present, all photons emitted above the Lyman-$\al$ frequency and redshifted below it in their passage through intergalactic space, would be absorbed, leading to a trough in the spectrum,  see~\cite{1965ApJ...142.1633G,Fan:2001vx}.

\paragraph{Radiation matter equality at $z_{\rm eq}\simeq 2.4\times 10^4$ : }
The photon  energy $\propto 1/\la$ increases  with redshift. While the baryon and dark matter densities only increase by the reduction of the volume, $\rho_m\propto a^{-3} =(1+z)^3$, the radiation density behaves like $\rho_{r}\propto (1+z)^4$ and  dominates at redshifts above the equality redshift given by $z_{\rm eq}\simeq 2.4\times 10^4(\Om_mh^2)$.

\paragraph{Nucleosynthesis at $z_{\rm nuc}\simeq 3.2\times 10^8$ : }
At temperatures above $T_{\rm nuc} \simeq 0.08$MeV there were sufficiently many photons with energies above the deuterium binding energy of $E_D=2.2$MeV in the Universe to prevent deuterium from forming. Once the temperature dropped below $T_{\rm nuc}$ deuterium formed and most of it burned into He$^4$ leaving only traces of deuterium, He$^3$ and Li$^7$ in the Universe.
The abundance of He$^4$ is very sensitive to the expansion rate which at the time of
nucleosythesis, $t_{\rm nuc}\simeq 206$sec,  is dominated by the relativistic particles at $T\sim0.1$MeV. In the standard model of particle physics these are the photon and 3 species of relativistic neutrinos. The helium abundance of the Universe, $Y_{\rm He} \simeq 0.25$, was the first indication that there are really 3 (and not more) families  of particles with a light neutrino.  Much later, this has been confirmed with much better precision by measurements of the Z-boson decay width at the LEP accelerator at CERN. The deuterium and He$^3$ abundance on the other hand are very sensitive to the baryon density in the Universe and it is a big success of modern cosmology that this independent earlier `measurement' agrees so well with the result from CMB anisotropies which we discuss below.

\paragraph{Neutrino decoupling at $z_{\nu}\simeq 6\times 10^9$:} 
At temperature $T\simeq 1.4$MeV weak interactions freeze out. The mean free path of neutrinos becomes larger than the Hubble scale so that they are essentially free streaming. 
They conserve their distribution while the momenta are simply redshifted. This can be absorbed in a redshift of the  temperature, $T\propto (1+z)$. 
Later, at $T\simeq m_e =511$keV, when electrons and positrons annihilate, the CMB photons are heated by this energy release but the neutrinos are not. Therefore, after this event the neutrino temperature is somewhat lower than the photon temperature,
\be
T_\nu = \left(\frac{4}{11}\right)^{1/3}T_\ga \,.
\ee
At present, like the CMB there should be a neutrino background at a temperature of about $T_{\nu 0}\simeq 1.95$K. Even if neutrinos are massive, they are expected to have an extremely relativistic Fermi-Dirac distribution which has been modified since decoupling only by redshifting of the momenta. This background has not been observed directly until today. However, its effects on the Helium abundance and, especially on the CMB are well 
measured~\cite{Fields:2014uja,Archidiacono:2013dua,Sellentin:2014gaa}.
 
 There may also exist additional very weakly interacting light particles or sterile neutrinos which do not interact weakly and can only be generated by neutrino oscillations.  Depending on their mass, they  may or may not have thermalised in the past and they may actually be the dark matter~\cite{Canetti:2012vf}. 
  
\paragraph{The QCD and electroweak transitions at $z_{QCD}\sim10^{12}$ and $z_{EW}\sim 10^{15}$ : }
At earlier times, when the temperature was $T>T_{QCD}\simeq 100$MeV
quarks and gluons were free. Only when the temperature dropped below $T_{QCD}$ did they confine into hadrons. According to present lattice gauge theory calculations, this transition is not a true phase transition but only a cross-over~\cite{deForcrand:2014tha}. This, however, depends on the neutrino chemical potential which is not well constrained~\cite{Schwarz:2009ii}.

Before that, at $T>T_{EW}\simeq 200$GeV the $W^{\pm}$ and $Z$ bosons were massless and weak interactions were as strong as electromagnetic interactions. At $T_{EW}$ the electroweak symmetry was broken by the Higgs mechanism, the Higgs became massive and gave masses to the standard model particles coupling to it, especially to the $W^\pm$ and to the $Z$, so that the weak interactions became weak. Within the standard model, for a Higgs with $m_H \simeq 125$GeV, also this transition is simply a cross over.

At present, there are  no cosmological observations which represent a relic of these transitions. Hence we are not certain that the Universe ever reached these temperatures.

\paragraph{SUSY breaking, baryogenesis, leptogenesis : }
If there is supersymmetry (SUSY) it must be broken below a few TeV which would then have happened before  electroweak symmetry breaking and might have led to the formation of dark matter if the latter is a neutralino. It may also be that at TeV or higher energy scales the baryon asymmetry in the Universe which is of order 
$ (n_b-\bar n_b)/n_b \simeq 10^{-10}$ has formed either directly or over leptogenesis. All these particle physics processes need physics beyond the standard model of particle physics which is rather uncertain. The only indications we have that they took place is the existence of both, dark matter and the baryon asymmetry. It may however well be that we shall have to revise our understanding of their emergence. 

\paragraph{Inflation : }
As we have discussed in Section~\ref{s:infla}, it seems very probable that there was an early inflationary phase which has ended in reheating leading to a hot Big Bang with a radiation dominated universe. Such a phase would not only solve the horizon and flatness problem, but it also predicts a spectrum of scale invariant curvature fluctuations as it has been observed in the CMB. Actually, since inflation has to terminate eventually, $\ep>0$ is required and typical inflationary modes predict slightly red spectra, $n_s<1$. The fact that Planck finds (for the minimal 6-parameter model with $r=0$, no running, standard neutrino sector etc.)
$$  n_s =0.9603\pm  0.0073\,,$$
i.e. a red spectrum with a significance of more than 5 standard deviations can be considered as a great success for inflation. 
Since we have no clear observational signature of the very high temperature universe, we only know for certain that reheating happened well before nucleosynthesis, hence $T_{\rm rh}
\gtrsim 1$MeV.

Interestingly, many inflationary models lead to 'eternal inflation', i.e., some parts  
of spacetime are always inflating and only in isolated 'bubbles' inflation terminates and leads to a hot thermal universe. In combination with ideas from string theory, these 'bubble-universes' can correspond to  different vacuum states of string theory, leading not only to  bubbles with different particle content, different interactions and different bubble sizes, but even with different numbers of large spatial dimensions. Since string theory has a 'landscape' of about 
$10^{100}$ vacua~\cite{Bousso:2007er,Bousso:2007gp}, this can 'explain' the smallness of the observed non-vanishing cosmological constant by the simple fact that physicists cannot live in a Universe with a much larger cosmological constant.
Hence the multiverse~\cite{Guth:2007ng,Bousso:2007gp,Garriga:2008ks} picture can lead to a less arbitrary formulation of the 'anthropic principle'.  On the other hand, when adopting this picture, we give up the possibility to ever find an explanation, e.g., for the value of the fine structure constant, other than the anthropic principle. 

\paragraph{Alternatives to inflation : } Are we sure that inflation ever happened or might the very early phase of the Universe have been very different? At present there are several 'alternatives to inflation'. Most of them are bouncing Universes. This idea goes back to G. Lema\^\i tre~\cite{1933ASSB...53...51L}: It is assumed that 
the observed expanding Universe emerged from a collapsing phase which 'bounced' into expansion. Many such bounces can follow each other with ever increasing entropy and hence flatness~\cite{Durrer:1995mz}.

There are many possibilities how this may happen, all of them need either a closed Universe or modifications of General Relativity e.g. loop quantum gravity~\cite{Bojowald:2012xy,Linsefors:2013cd}, the pre-big bang model of string cosmology~\cite{Gasperini:2002bn} or the 'ekpyrotic' or 'cyclic' universe~\cite{Khoury:2003rt}. If expansion follows from a long contracting phase, clearly the horizon problem is solved. Via uncertain modifications of General Relativity, these models also avoid the singularity problem. The contracting universe, which within General Relativity usually leads to a big crunch singularity, stops at some very high density, where corrections become relevant, and turns into expansion. Despite the untested but often well motivated modifications of General Relativity, these models cannot solve the flatness problem. They usually just assume a homogeneous and isotropic universe. Actually, small initial density fluctuations grow exponentially during a contracting phase. Furthermore, it is not easy to obtain a nearly scale invariant spectrum of initial fluctuations in these models. One interesting consequence, however, is that such models usually predict a negligible tensor to scalar ratio. More precisely, the tensor spectrum is very blue and has nearly no power on the large scales which are tested with CMB experiments. Therefore, the discovery of a scale invariant spectrum of tensor fluctuations, i.e., gravitational waves would rule out most of these models.

\subsection{Cosmological parameters}

Let us define the density parameter of some component $X$  with energy density $\rho_X$  by
$$ \Om_X  =  \frac{8\pi G\rho_X(t_0)}{3H_0^2} \quad \mbox{ or } ~ \om_X = h^2\Om_X  =  \frac{8\pi G\rho_X(t_0)}{3(H_0/h)^2} \,.$$
The second quantity has the advantage that it is proportional to the energy density of the component $X$ via a well known numerical constant while in the first, the significant uncertainty of $H^2_0$ enters the value of $\Om_X$.
We consider the radiation density $\rho_r(t_0) \propto \om_r$ as fixed, since we know both the photon and neutrino temperatures with very high accuracy (even though we have not measured the neutrino background, we can infer its temperature theoretically as $T_\nu=(4/11)^{1/3}T_{\mbox{\tiny CMB}}$, see, e.g.~\cite{Weinberg:2008}). Then $\om_m=\Om_mh^2$ determines matter and radiation equality and thereby the wave numbers of fluctuations which enter the Hubble scale  still during the radiation dominated era. Curvature fluctuations with these wave numbers decay after horizon entry until equal matter and radiation, while curvature fluctuations  entering the Hubble scale during the matter dominated regime, always  remain constant.

\begin{figure}
\begin{center}
\includegraphics[width=6.5cm]{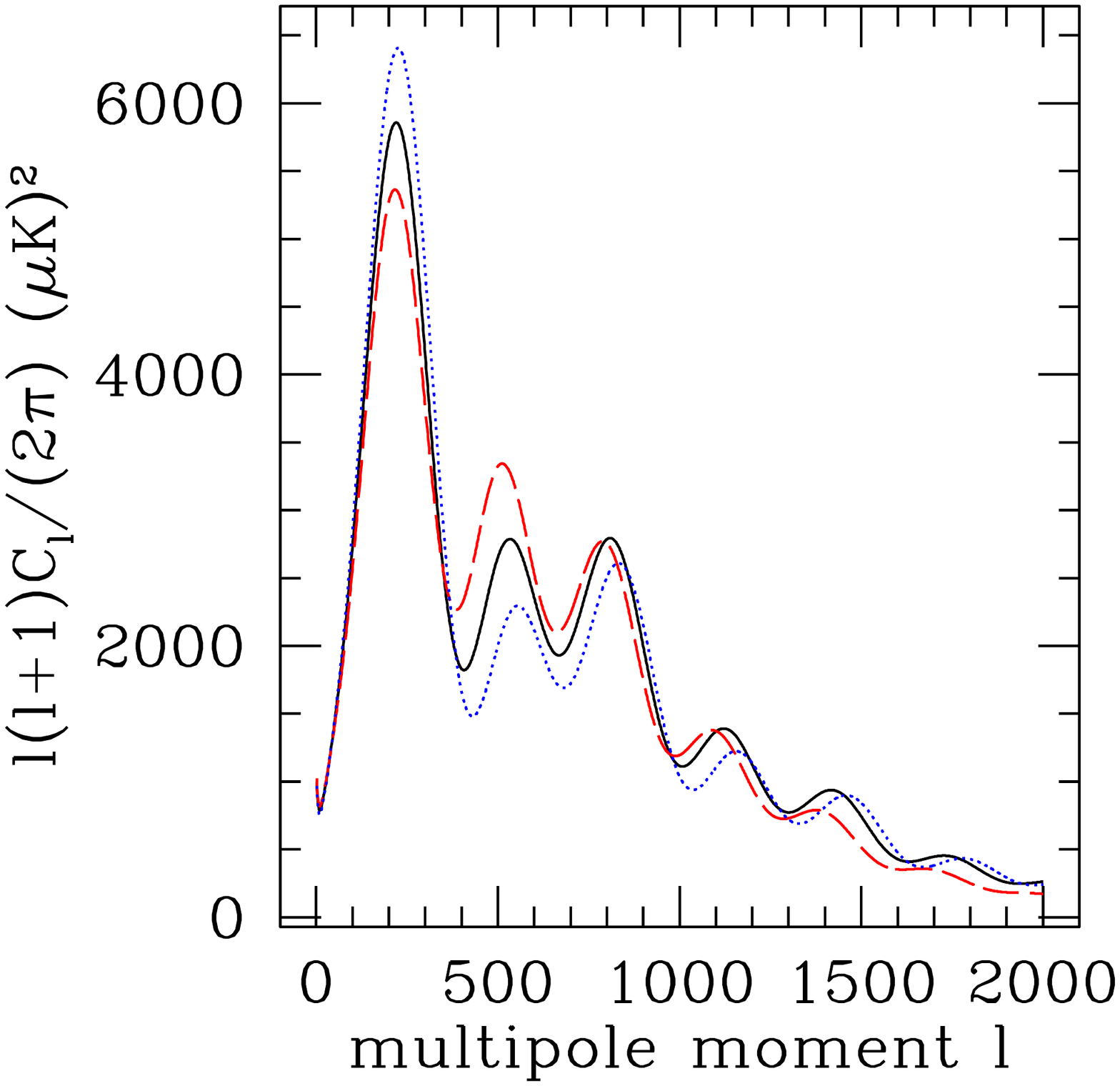}\qquad
\includegraphics[width=6.5cm]{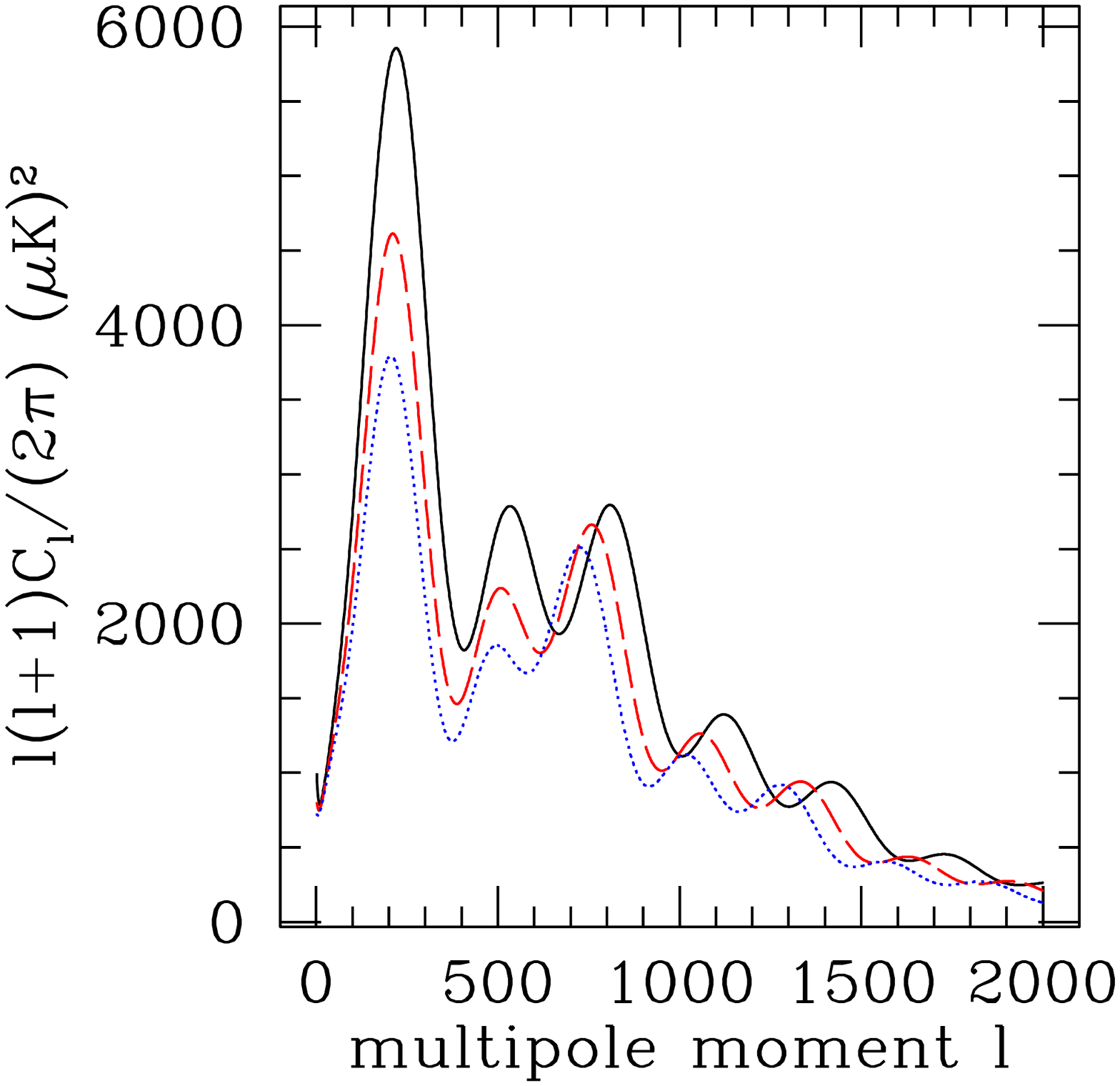}\end{center}
\caption{\label{f:omb} The CMB anisotropy spectrum for
$\om_b = 0.02$ (solid line, black),  $\om_b = 0.03$ (dotted, blue) and
$\om_b = 0.01$ (dashed, red) is shown in the left panel. Note that the asymmetry of even and odd peaks is enhanced if the baryon density is increased. On the right hand panel $\om_b=0.02$
is fixed and three different values for the matter density are
chosen, $\om_m=0.12$ (solid, black),  $\om_m=0.2$ (dashed, red)
$\om_m=0.3$ (dotted, blue). Higher values of $\om_m$
also lead to a stronger peak asymmetry. Smaller
value of  $\om_m$ boosts the height especially of the first
peak. A detailed discussion of the parameter dependence is given in Ref.~\cite{2008cmb.book}.}
\end{figure}

The transfer function also depends on the baryon density proportional to $\om_b=\Om_bh^2$ in multiple ways:
First of all, before photon decoupling the baryon-photon fluid performs the acoustic oscillations mentioned above.  Without baryons the amplitude of these fluctuations is constant.
The presence of baryons leads to an amplification of the compression peaks (over densities) and a reduction of the expansion peaks (under densities) by gravitational attraction. Furthermore, baryons slightly reduce the sound speed of the baryon-photon fluid as they contribute to the energy density but not to the pressure. Once photons decouple their mean free path grows and the photon fluctuations on small scales are damped by diffusion. In cosmology this is called Silk damping~\cite{1968ApJ...151..459S}.  Photons diffuse from over densities into under densities. The details of this decoupling process and especially how fast it takes place also depend on the baryon density.
The dependence of CMB anisotropies on the baryon and matter densities is shown in Fig.~\ref{f:omb}. 

The CMB spectrum of course also depends on the initial conditions which for purely scalar curvature perturbations are given by $\De_{\RR}$ and $n_s$.

Finally, the angle onto which a given wavelength in the CMB sky is projected depends on the distance from us to the last scattering surface which is strongly affected by the matter content of the Universe. In a Universe with radiation, matter, curvature  and a cosmological constant,  we have
\bea
\hspace*{-2.5cm} d_A(z_*) &\hspace*{-0.8cm}=& 
\frac{1}{(z+1)}\chi\left(\int_0^{z_*}
\frac{dz}{H(z)}\right)   \nonumber \\  && \hspace*{-0.8cm} =~ \frac{1}{(z+1)}\chi\left(\int_0^{z_*}
\frac{dz}{H_0\sqrt{\Om_r (z+1)^4 +\Om_m (z+1)^3 +
\Om_\La+\Om_K (z+1)^2 }}\right)\,. 
\eea
Here $\Om_r,~\Om_m,~\Om_K$ and $\Om_\La$ are the present density parameters
of the different components  so that $\Om_m+\Om_r+\Om_K+\Om_\La=1$ and $z_*$ is the redshift of decoupling, i.e. the redshift at which the CMB photons are emitted.
If dark energy is not simply a cosmological constant, but evolving, we have to replace $\Om_\La$ by $\Om_{\rm de}(z)$. With the normalization such that  the present value of the scale factor is unity we have $\Om_K = -K/H_0^2$.
The function $\chi(r  )$ is given by
$$ \chi(r  ) =\left\{\begin{array}{lcc}  
\frac{1}{\sqrt{K}}\sin\left(\sqrt{K}r\right) & \mbox{if} & K>0\\
\frac{1}{\sqrt{-K}}\sinh\left(\sqrt{-K}r\right) & \mbox{if} & K<0\\
r & \mbox{if} & K=0 \,.\end{array}
 \right.
$$
Clearly, this distance strongly depends on curvature, on $H_0$ and also on $\Om_\La$.
The angular extension of the wavelength corresponding to the first peak in the sky is given by
\be
\theta_* = \frac{r_s}{d_A(z_*)} \,,
\ee
where $r_s$ is the sound horizon at the last scattering surface given by
\be
r_s=a(z_*)\int_0^{t_*}\frac{c_sdt}{a(t)} = \frac{2}{\sqrt{3r\om_m}}
\log\left(\frac{\sqrt{1+z_* +R} +\sqrt{\frac{(1+z_*)
R\omega_r}{\omega_m} +R}}{\sqrt{1+z_*}\left(1 +
 \sqrt{\frac{R\omega_r}{\om_m}}\right)}\right) \,, \qquad R= \frac{3\om_b}{4\om_\ga} \,.
\ee
Here $\om_r=\Om_rh^2$
is the radiation density including neutrinos and $\om_\ga=\Om_\ga h^2$ is the photon density.

\subsection{Precision cosmology}\label{ss:prec}
Already in the 90ties it was realized that with a precise determination of the CMB anisotropies and its comparison with calculations we can estimate cosmological parameters to an unprecedented precision~\cite{Bond:1993fb}. Soon, fast codes to calculate CMB anisotropies to first order in perturbation theory were developed~\cite{Seljak:1996is,Lewis:1999bs}. Efficient Markov Chain Monte Carlo (MCMC) routines to search for the best fit in a mul\-ti\-dimen\-sional cosmological parameter space 
followed~\cite{Lewis:2002ah,Lesgourgues:2011re,Audren:2012wb}.
At present these codes are used to estimate cosmological parameters from CMB anisotropies. They announce a numerical precision of about 0.1\% for cosmological parameters which are not too far away from the standard values.

Below I shall only present the findings for the standard $\La$CDM model from present CMB data. But what is especially important is that there is no other simple model which cannot
be continuously deformed into $\La$CDM which fits all the data. As an example, let me mention the DGP model~\cite{Dvali:2000hr}, a 5d-braneworld model of the Universe leading also to accelerated expansion at late time which has been a rival to  $\La$CDM but is now excluded combining WMAP with other cosmological data~\cite{Xu:2010um}.  Other attempts are massive gravity~\cite{deRham:2010kj} or bigravity~\cite{Fasiello:2013woa} which either are not fully worked out or are ruled outl~\cite{deRham:2014zqa,Cusin:2014psa} as well as quintessence models~\cite{Caldwell:1997ii} which can be continuously deformed to  $\La$CDM.  An interesting exception is a recently proposed non-local model of massive gravity~\cite{Dirian:2014bma}.

Apart from linear perturbation theory discussed in Section~\ref{ss:pert}, the following two additional physical effects are included in the calculation of CMB anisotropies: 

1) \emph{Lensing:} Due to lensing by foreground inhomogeneities photons are deflected and we see them not exactly in the direction into which they have been emitted. Since the fluctuations are already first order this lensing effect is of second order, but it is nevertheless  relevant, see~\cite{Lewis:2006fu} for a review. On small angular scales, $\ell \gtrsim 1000$ it changes the resulting spectra by 10\% and more. Looking in direction $\bn$, we actually see the temperature fluctuation not as it was at position $\bn r_*$ but at position $(\bn+\bm{\al})r_*$, where $\bm\al$ denotes the deflection angle. To first order in perturbation theory the deflection angle is given by
\be
\bm\al = -2\int_{0}^{r_*}dr\frac{\chi(r_*- r )}{\chi(r_*)\chi(r )}
  \bm{\nabla}_{\perp}\Psi (t(r ) ,r,\vth,\vph) \equiv  \bm{\nabla}_{\perp}\phi (\vth,\vph)~.
\ee
Here $\Psi$ is the gravitational potential (the Bardeen potential), $\bn$ is given by $(\vth,\vph)$ and $\bm{\nabla}_{\perp}$ is the gradient on the sphere of photon directions.

$$ 
\phi (\vth,\vph) =-2\int_{0}^{r_*}dr\frac{\chi(r_*- r )}{\chi(r_*)\chi(r )}\Psi (t(r ),r,\vth,\vph) 
$$
is the lensing potential (see contribution by Cliff Will~\cite{Will:2015} for the history and  applications of relativistic light deflection).
For typical lines of sight, this deflection angle from the CMB is several arc minutes.

As we shall see in the next sections, lensing is also relevant for polarization and, since it is a second order effect, it introduces non-Gaussianities.

2)  \emph{Reionization:}  As discussed above, at some redshift of order $z_{\rm ri}\sim 10$ the hydrogen in the Universe is re-ionized by the UV light from the first stars. Since this process is complicated and cannot be calculated reliably, it is taken into account in the CMB codes by an effective optical depth due to reionisation, $\tau_{\rm ri}$ or simply by a fixed reionisation redshift,  $z_{\rm ri}$. In the calculations $\tau_{\rm ri}$ is treated as an additional unknown parameter to be fitted by the data.
\vspace{0.15cm}

In addition to the temperature anisotropy spectrum and the polarization spectrum discussed in the next section, the Planck satellite experiment has extracted the spectrum of the lensing  potential with, however, still modest accuracy.

\begin{figure}[ht]
\begin{center}
\includegraphics[width=8cm]{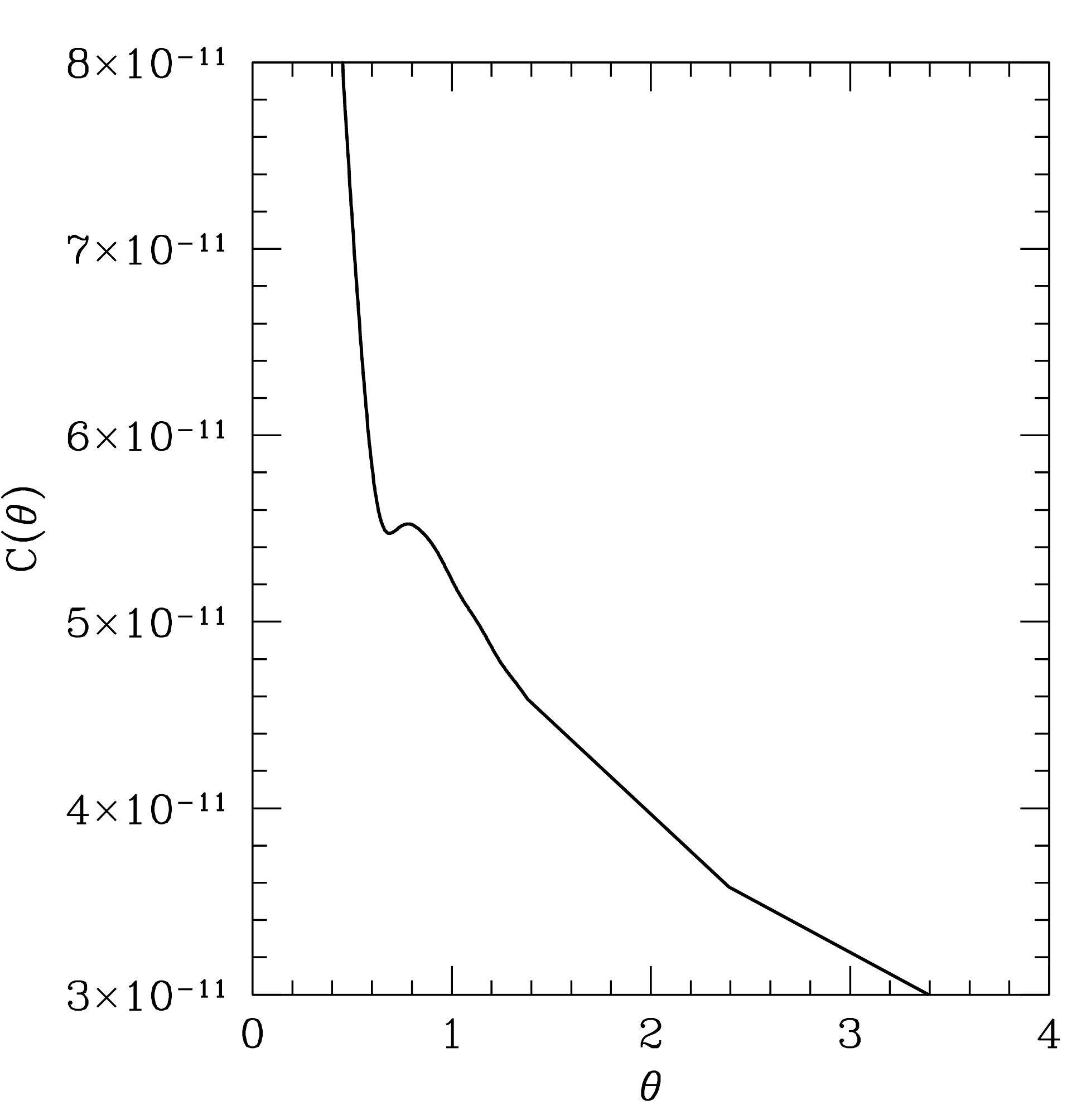}
\end{center}
\caption{\label{f:correl}The CMB correlation function for typical values of the cosmological parameters in degrees. The most prominent feature is the acoustic sound horizon which is seen at $\theta\simeq0.6^o$.}
\end{figure}

The quantity which is best determined by the CMB anisotropies is the angle subtended by the sound horizon at last scattering~\cite{Ade:2013zuv},   
$$ \theta_* = \frac{r_s}{d_A(z_*)} = (1.04131 \pm 0.00062)\times 10^{-2} \simeq 0.6^o  \,.
$$
This is a very prominent feature not only in the power spectrum but also in the correlation function as shown in Fig.~\ref{f:correl}.

\begin{figure}[ht]
\begin{center}
\includegraphics[width=13cm]{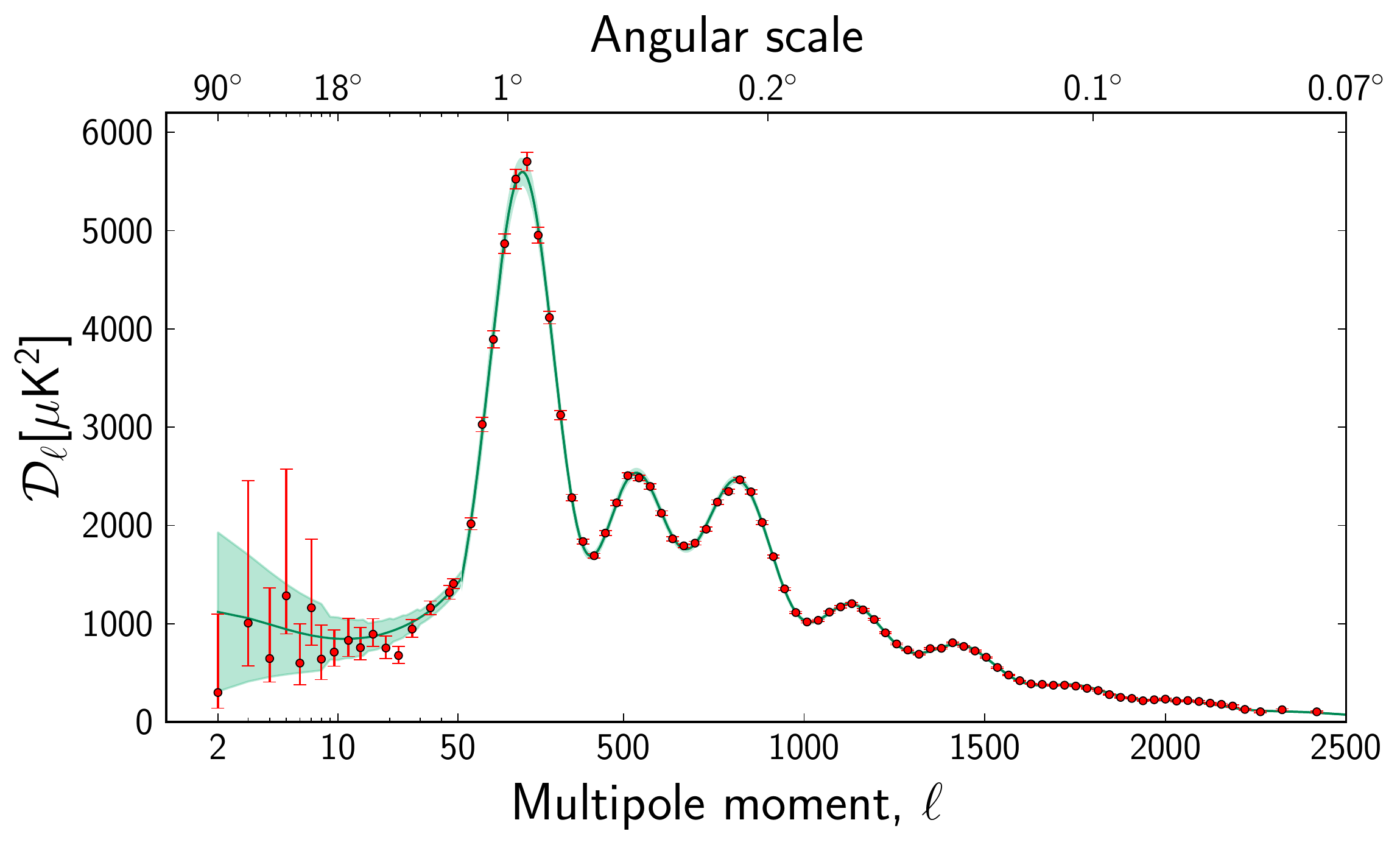}
\end{center}
\caption{\label{f:Planck-spec}The CMB power spectrum as seen by 
Planck~\cite{Ade:2013kta}. The red dots with error bars are the data points and the green line is the best fit theoretical model. The shaded region indicates the theoretical error from cosmic variance. The precision is essentially cosmic variance limited out to $\ell\simeq 2000$.}
\end{figure}

The power spectrum of pure scalar perturbations in a Universe with vanishing background curvature is an excellent fit to present data as can be seen in Fig.~\ref{f:Planck-spec} where we compare the best fit calculated spectrum for $K=0$ and $r=0$ with the data.

The shaded region in the figure indicates 'cosmic variance', i.e., the statistical error due to the fact that we have only one sky at our disposal and therefore to obtain e.g. $C_2$ we can average at best over five values $a_{2 m}$. For Gaussian fluctuations this leads to a cosmic variance of 
$$ \frac{\De C_\ell}{C_\ell} = \sqrt{\frac{2}{2\ell +1}} \,. $$
In practice, since the region close to the galactic plane cannot be used for CMB analysis this error is increased to  $\sqrt{\frac{2}{(2\ell +1)f}}$, where $f$ denotes the  fraction of the sky used for the analysis. For a satellite experiment this is typically $f\simeq 0.7$.

The best fit parameters from the Planck analysis~\cite{Ade:2013zuv}, including the polarization of the WMAP experiment, are given in Table~\ref{t:param}.

\begin{table}
\begin{center}
\begin{tabular}{||c|c||}
\hline   &  \\  parameter & value \\[3pt] \hline & \\
$\om_b$ & $0.02205\pm 0.00028$\\[3pt]
$\om_c$ & $0.1199 \pm 0.0027$\\[3pt]
$\theta_*$& $(1.04131\pm 0.00063)\times 10^{-2}$\\[3pt]
$n_s$ & $0.9603\pm 0.0073$\\[3pt]
$\log(10^{10}\De_R)$ & $3.089^{+0.024}_{-0.027}$\\[3pt]
$\tau_{\rm ri}$ & $0.089^{+0.012}_{-0.014}$\\[3pt]
\hline
\end{tabular}
\end{center}
\caption{\label{t:param} The Planck~\cite{Ade:2013zuv} parameters for the best fit
model with vanishing curvature and purely scalar perturbations. Here $\om_c =\Om_ch^2$ is the density parameter of cold dark matter so that the total matter density parameter is $\om_m=\om_c+\om_b$.}
\end{table}
Interestingly, for a flat model, $\Om_K=0$ hence $h= \sqrt{\om_m/(1-\Om_\La)}$, the distance, $d_A(z_*) = r_s(\om_m,\om_b)/\theta_*$ leads to a Hubble parameter of $h=0.673\pm0.012$ and $\Om_\La =0.685\pm 0.017$. This Hubble parameter is nearly 2.5 standard deviations smaller than the one inferred by local observations which yield~\cite{Riess:2011} $h=0.738\pm 0.024$.

\begin{figure}[ht]
\begin{center}
\includegraphics[width=7.5cm]{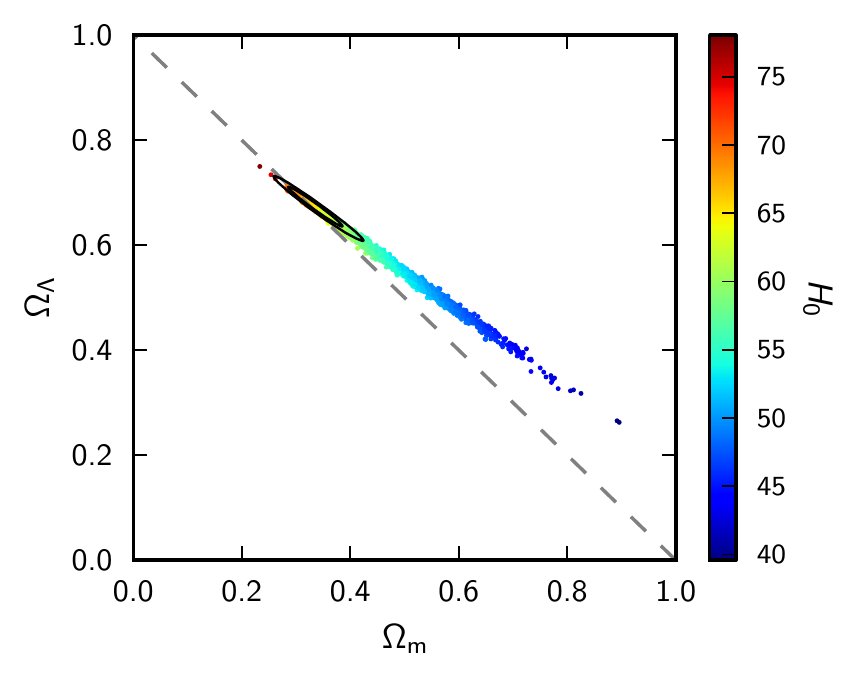} \quad
\includegraphics[width=7.5cm]{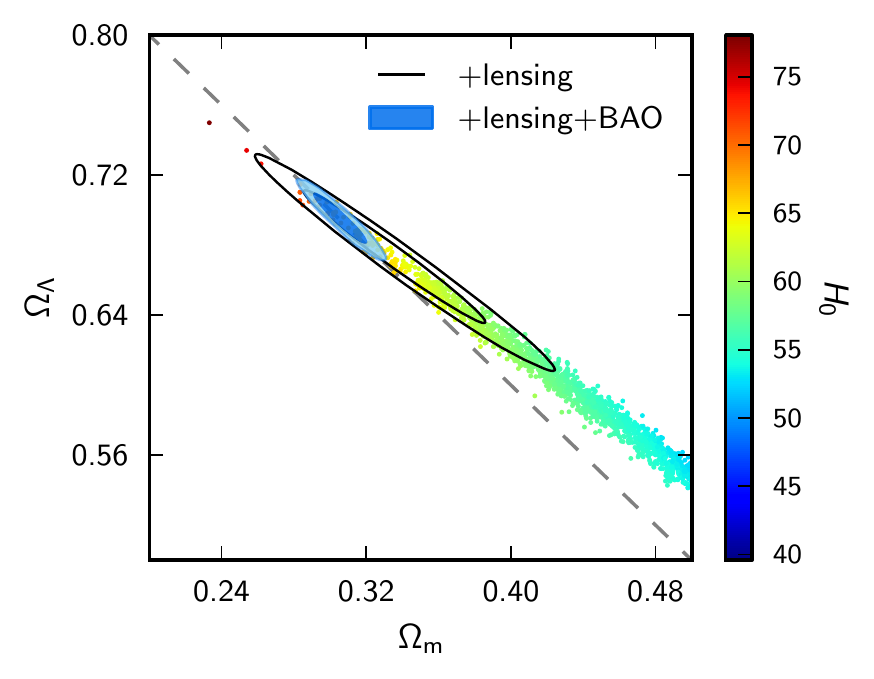}
\end{center}
\caption{\label{f:degen} The degeneracy between $\Om_\La$ and $H_0$ is shown (left panel). Once CMB lensing, and BAO's are included the degeneracy is lifted. Figure from~\cite{Ade:2013zuv}.}
\end{figure}

The CMB is very sensitive to $\om_b$, $\om_m$ and of course to the primordial power spectrum, characterized in the simplest case by $n_s$ and $\De_{\RR}$. However,
CMB spectra from cosmologies with the same matter densities, the same primordial power spectra and the same area distance $d_A(z_*)$ are nearly identical. Therefore, the CMB measures $h$, $\Om_K$ and $\Om_\La$ mainly via their contribution in the area distance to the last scattering surface. Since the Friedmann equation gives one relation between these parameters, e.g., $\Om_K=1-\Om_\La -\om_m/h^2$, this leaves us with a degeneracy between e.g. $h$ and $\Om_\La$. This geometrical degeneracy is lifted somewhat by the late integrated Sachs-Wolfe effect, see Eq.~(\ref{e:dTint}), which is sensitive mainly to $\Om_\La$   and more prominently by lensing. Lensing depends differently on both $h$ and $\Om_\La$ than $d_A(z_*)$ and therefore, together with the accurate determination of $d_A(z_*)$, allows us to determine both parameters. The degeneracy is completely lifted if we combine the data also with baryon acoustic oscillations (BAO's).  These generate the same acoustic peaks which we see in the CMB but now in the matter power spectrum which we measure at much lower redshifts. Since we can determine the BAO's at different redshifts, this also allows us to break the degeneracy, see~Fig.~\ref{f:degen} right panel. 

Combining CMB temperature anisotropy and lensing with BAO data, Planck can set a very stringent limit on curvature,
\be\label{e:OmK} \Om_K =-0.0005\pm0.0066   \quad  \mbox{at }~ 95\% \mbox{ confidence.}\ee

Also, fitting the temperature anisotropies as a combination of scalar and tensor contributions, Planck can derive a limit on the tensor to scalar ratio,
\be\label{e:r}  r  \le 0.11    \quad   \mbox{at }~  95\%\mbox{ confidence.} \ee
For this limit it is assumed that the scalar spectral index $n_s$ is constant, i.e. independent of scale, `no running'. When running is admitted the limit degrades to $r<0.26$.

\subsection{Non-Gaussianities}
Curvature and gravitational wave perturbations generated during inflation  typically obey Gaussian statistics. Within linear perturbation theory this leads to Gaussian CMB temperature anisotropies and polarization. The amplitude of non-Gaussianities from typical slow roll inflationary models are of the order of the slow roll parameters and hence very small~\cite{Maldacena:2002vr}. There are, however well motivated inflationary models which predict appreciable non-Gaussianities. Furthermore, since the square of a Gaussian field is not Gaussian, non-linearities of gravity also induce non-Gaussianities. In the analysis of the Planck experiment this effect has been used in the determination of the lensing power spectrum from the temperature anisotropy data~\cite{Ade:2013tyw}.

A simple characterisation of non-Gaussianities is the bispectrum, the Fourier transform of the three point function. For the Bardeen potential $\Psi$ we set
\be
\langle\Psi(\bk_1)\Psi(\bk_2)\Psi(\bk_3)\rangle = (2\pi)^3\de(\bk_1+\bk_2+\bk_3)B(k_1,k_2,k_3) \,.
\ee
Let us first consider non-Gaussianity  of the so called 'local' type given by  
$$\Psi(\bx) = \Psi_G(\bx) + f^\text{(loc)}_{NL}\left(\Psi_G^2(\bx) -\langle\Psi_G^2\rangle\right) \,,$$
where $\Psi_G$ is a Gaussian field.
The bispectrum then becomes
\be
B^\text{(loc)}(\bk_1,\bk_2,\bk_3 ) =2f^\text{(loc)}_{NL} \left(\sum_{\rm perm}P(k_1)P(k_2)  \right)\,,
\ee
where $P(k)$ denotes the power spectrum of $\Psi$.
Here the sum is over the three permutations of the wave numbers, (1,2), (1,3) and (2,3).
For typical non-Gaussianities, e.g. coming from second order perturbation theory, this parameter is scale independent. It is however very sensitive to the shape of the triangle formed by $(\bk_1,\bk_2,\bk_3)$ which characterizes the type of non-Gaussianity.
For simple, local quadratic non-Gaussianities $f^\text{(loc)}_{NL}$ is dominant in the 'squeezed limit' i.e., when one of the sizes of the triangle  formed by $(\bk_1,\bk_2,\bk_3)$ tends to zero. The equilateral respectively orthogonal bispectra  dominate when the triangles formed by the $\bk_i$ vectors are equilateral respectively orthogonal. They are of the form
\bea
 \hspace*{-2.4cm} B^\text{(equi)}(\bk_1,\bk_2,\bk_3 ) =&& \nonumber\\  \hspace*{-2.2cm}
 6f^\text{(equi)}_{NL} \left[\sum_{\rm perm}(P(k_1))^{1/3}(P(k_2))^{2/3}P(k_3) -2(P(k_1)P(k_2)P(k_3))^{2/3}- \sum_{\rm perm}P(k_1)P(k_2)   \right] , 
 && \nonumber\\
 \hspace*{-2.4cm} B^\text{(ort)}(\bk_1,\bk_2,\bk_3 ) =&& \nonumber\\  \hspace*{-2.2cm}
 18f^\text{(ort)}_{NL} \left[\sum_{\rm perm}(P(k_1))^{1/3}(P(k_2))^{2/3}P(k_3) -\frac{8}{3}(P(k_1)P(k_2)P(k_3))^{2/3}- \sum_{\rm perm}P(k_1)P(k_2)  \right] .
 &&\nonumber
\eea
These forms of the bispectrum have been obtained in different models of inflation, see for example~\cite{Creminelli:2005hu,Senatore:2009gt}.
Planck has published limits on  $f_{NL}$ for three different shapes which approximate the ones given above.
To arrive at them, the significant lensing contribution had to be subtracted~\cite{Ade:2013ydc}.
\bea
f_{NL}^{(X)}=\left\{\begin{array}{ll}
2.7 \pm 5.8 & X=\mbox{ loc} \\  -42\pm 75 & X= \mbox{ equi} \\   -25\pm 39  &  X=\mbox{ ort} \end{array}  \right.
\eea
The errors given correspond to 68\% likelihood. 
It seems as if the local shape would be much better constrained than the equilateral or orthogonal ones, but this is mainly a consequence of the definition of $f^{(X)}_{NL}$'s.  
Clearly, there is no evidence for primordial non-Gaussianity in the present CMB data.

Non-Gaussianities are of course not given by the bispectrum alone. They may lead to a vanishing bispectrum e.g. for symmetry reasons but non-vanishing reduced four point function, i.e.,  trispectrum or any other reduced higher moments which are absent in a Gaussian distribution.
Apart from looking for higher moments there are also other techniques to find the non-Gaussianity of fluctuations like, e.g., by analyzing void statistics or simply the shape of the 1-point distribution function.
 
 It has to be noted, however, that in cosmology measuring non-Gaussianity is always intimately related to statistical isotropy (and homogeneity). When we determine the distribution of e.g. the mean temperature fluctuation on the angular scale of one degree, we cannot take an ensemble average, but we just average over all possible directions in the sky, assuming that the fluctuations are statistically isotropic so that this is a good approximation to an ensemble average. If we find that the distribution of these fluctuations is not Gaussian but, e.g., bimodal, this may signify two things: either the CMB fluctuations are indeed non-Gaussian or the mean amplitude is different in one part of the sky than in another, i.e.,  there is a preferred direction and the Universe is not statistically isotropic. This example shows that the two intrinsically independent properties of statistical isotropy and Gaussianity cannot be tested independently since we can observe only one CMB sky.
 
\section{CMB polarization}\label{s:pol}
\begin{figure}[ht]
\begin{center}
 \includegraphics[width=0.5\linewidth]{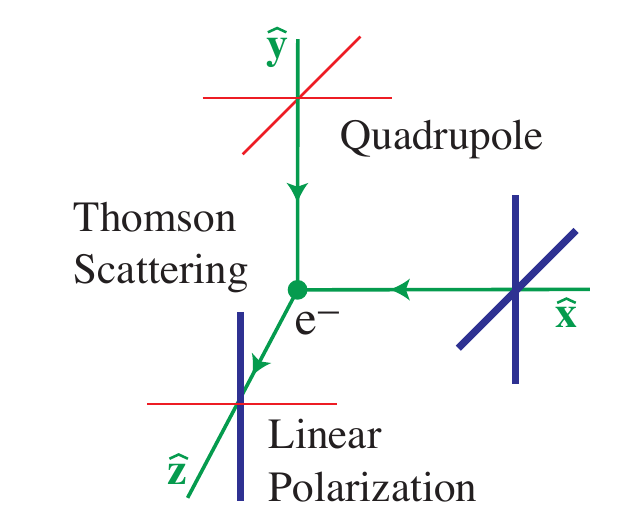}
 \end{center}
\caption{\label{f:pol} The Thomson cross section depends on polarization. Scattering by an angle $\theta$ is suppressed by a factor $\cos^2\theta$ if the polarization vector lies in the scattering plane.  In the depicted situation with $\theta=\pi/2$, the photon with blue polarization directions is  scattered only if its polarization is vertical while the photon with red polarization directions is  scattered onl if its polarization is horizontal.  A quadrupole anisotropy in the (unpolarized) incoming radiation intensity seen by the scattering electron generates a net polarization of the outgoing radiation. Figure from~\cite{Hu:2001bc}.}
\end{figure}

Thomson scattering is not isotropic. The probability of scattering a photon with a polarization vector in the scattering plane is suppressed by a factor $\cos^2\theta$, where $\theta$ is the angle between the direction of the incoming and the outgoing photon. This factor ensures that no 'longitudinal' photons are generated by Thomson scattering. If the radiation intensity as seen from the scattering electron has a non-vanishing quadrupole anisotropy, this leads to a net polarization of the outgoing radiation, as depicted in Fig.~\ref{f:pol}. 

This polarization is generated on the last scattering surface and to some small extent again when the Universe is re-ionized.  Within linear perturbation theory, the polarization pattern from scalar perturbations is always in the form of a gradient field on the sphere, called $E$-polarization, while the polarization induced by gravitational waves has both a gradient ($E$) 
and a curl component. The latter is called $B$ polarization. The detection of $B$-polarization would therefore be a unique signal of tensor modes. Unfortunately the situation is not so clear-cut as non-linearities in the evolution also lead to $B$-polarization. Especially, lensing of scalar $E$-modes induces $B$-polarization. Therefore, if the tensor to scalar ratio is too small, it is very difficult to ever detect tensor modes.

\begin{figure}[ht]
\begin{center}  
 \includegraphics[width=0.45\linewidth]{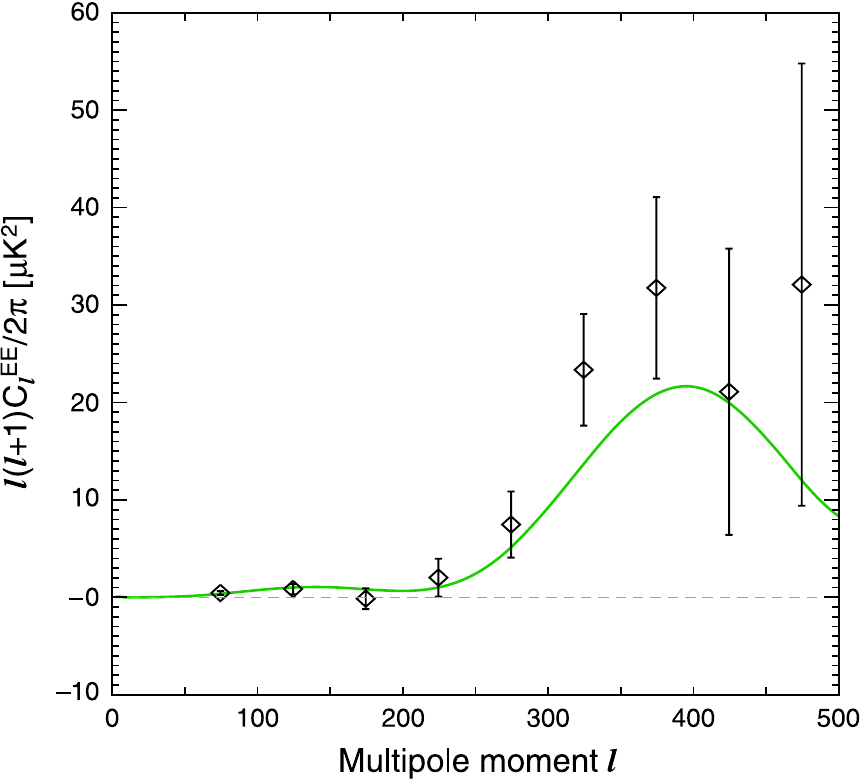}\qquad   \includegraphics[width=0.45\linewidth]{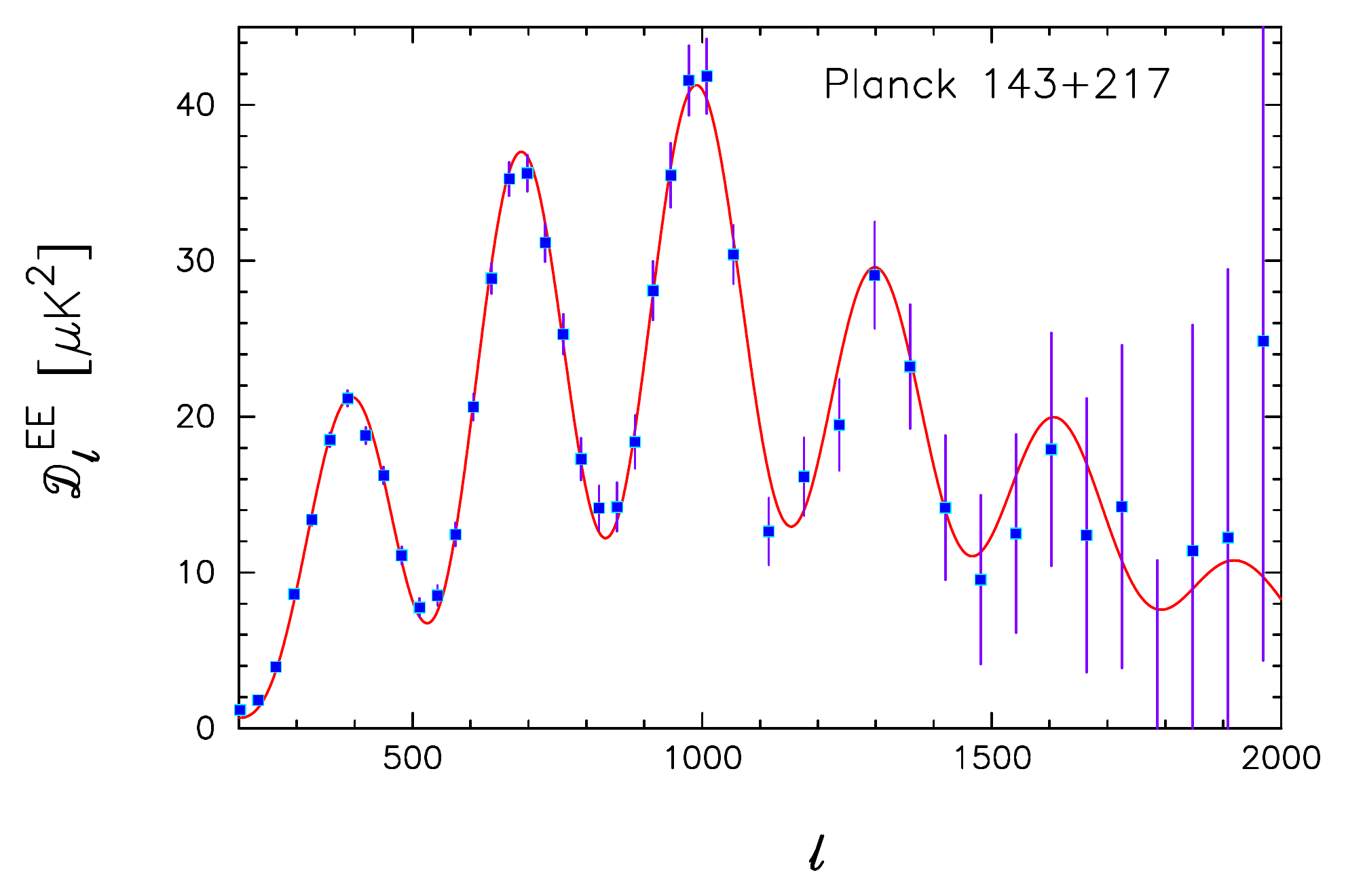}
 \end{center}
\caption{\label{f:polWMAP} In the left panel the WMAP9 polarization spectrum is shown. On the right a preliminary Planck polarization spectrum from the 143GHz and 217GHz channels is shown. The predicted polarization spectrum for the best fit mode inferred from the temperature anisotropy data is shown as solid red line. Figures from~\cite{2011ApJS..192...16L} and~\cite{Ade:2013ktc}.}
\end{figure}

\begin{figure}[ht]
\begin{center}  
 \includegraphics[width=0.75\linewidth]{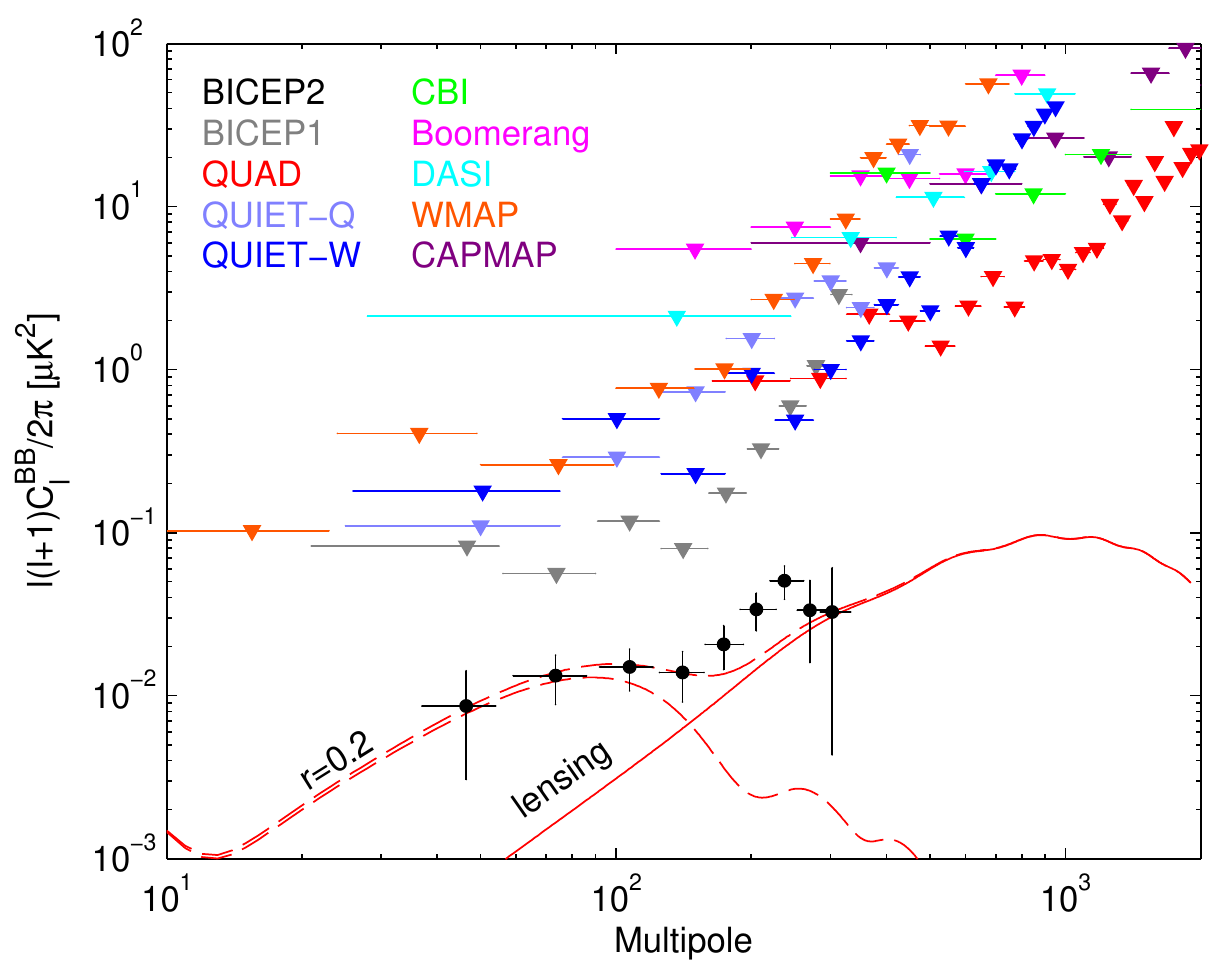}
 \end{center}
 \caption{\label{f:Bicep2} The B-mode polarization measurements at the time of the
 BICEP2 publication. Apart from the BICEP2 results (black), these are all upper limits. Also indicated are the theoretical lensed E-modes from scalar perturbations (solid line) and the theoretical tensor  
spectrum of B-modes for $r=0.2$ (dashed line).  Figure from~\cite{Ade:2014xna}.}
 \end{figure}

The best published polarization data today is the WMAP and Planck data shown in Fig.~\ref{f:polWMAP}. It is compatible with pure E-polarisation.
Earlier this year, the BICEP2 experiment~\cite{Ade:2014xna} announced the detection of a B-polarization signal with an amplitude leading to a tensor to scalar ratio of $r=0.2$, see Fig.~\ref{f:Bicep2}. 
 
This has stirred a tremendous excitement in the community as such a large tensor to scalar ratio requires an inflationary energy scale of about $E_\text{inf}=V^{1/4} \simeq 2\times 10^{16}$GeV.
This would first of all tell us that in the CMB anisotropy and polarization we find information
on the physics at this very high energy scale, more than 12 orders of magnitude higher than the highest energy achieved in a particle physics accelerator, namely in the LHC at CERN.
Furthermore, it would indicate that the inflaton field has rolled down by several Planck energies during inflation~\cite{Lyth:1996im}.  This might be an indication that quantum gravity effects are relevant for inflation and therefore inflation might be a portal towards observations of quantum gravity.

Soon after these results were published, several researchers criticised them as possibly due to dust.  Since the BICEP2 data come from only one frequency, they rely on other datasets, especially Planck, to estimate the dust contribution in their data. Recently, the Planck team together with the BICEP and Keck teams  have
reanalysed the data using the detailed dust measurements which are possible with the large frequency coverage of Planck~\cite{Ade:2015tva}. They concluded that the BICEP findings are compatibly with purely dust and just yield an upper limits for the tensor to scalar ratio of
 $r<0.12$.

\section{The future}\label{s:futur}
In principle all future research on cosmology is affected by the discovery of the CMB.
In this section I describe some of the directions of research which are most strongly influenced by it either because they are  concerned by the CMB itself or because they represent a natural extension of the CMB studies.

\subsection{B-polarization and tensor modes}
The Planck satellite has measured the temperature fluctuations with cosmic variance limited error bars down to scales of a few arc minutes, where foregrounds start to dominate. Therefore we do not expect much further information on the CMB from temperature measurements.
However, as we have seen above, B-polarization has not yet been discovered.
A value of $0.1>r>0.001$, has tremendous implications for cosmology: It fixes the inflationary scale at roughly the GUT (grand unified theory) scale, the scale at which the coupling strengths of electromagnetic, weak and strong interactions unify.
Furthermore, for such a large tensor to scalar ratio, inflation must be of the `large field' type where the inflaton field evolves over several Planck scales during inflation.  In this case it is hard to understand why an effective field theory calculation can make sense. In the context of effective field theories one supposes that  the inflaton is an effective  `low energy' degree of freedom of a more complicated theory at higher energy the potential of which is given in the form
\be 
V(\phi) = \frac{m^2}{2}\phi^2 + \sum_{n=2}^N\la_n\frac{\phi^{2n}}{m_P^{2n-4}} \,. 
\ee
The higher dimensional, Planck-mass suppressed operators of the form $\phi^{2n}/m_P^{2n-4}$ cannot be suppressed if $\phi$ varies over a range larger than $m_P$. Hence we have to rethink the effective field theory approach to inflation.

In other words, whenever $r$ is big enough to be measurable, its detection will be of uttermost importance not only for cosmology but for all of high energy physics. 

\begin{figure}[ht]
\begin{center} 
 \includegraphics[width=0.8\linewidth]{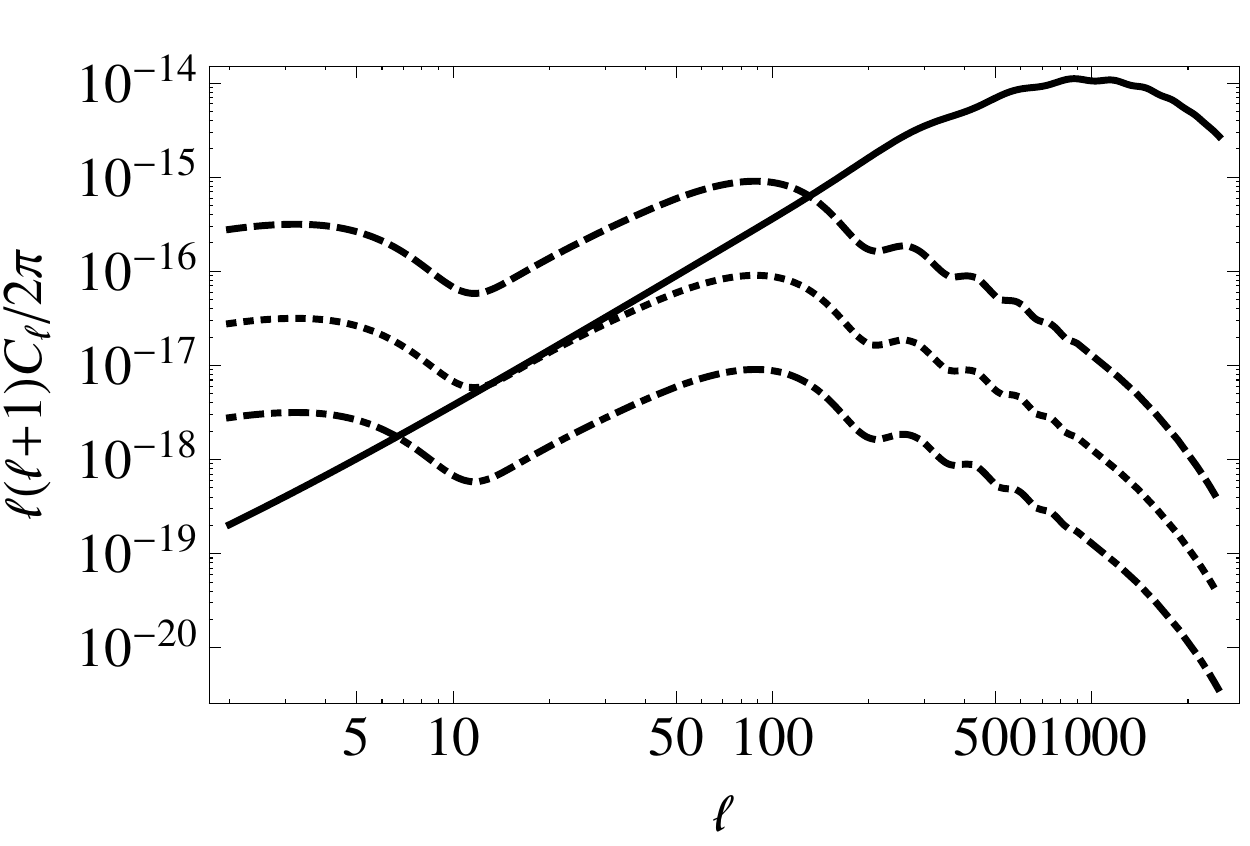}
 \end{center}
\caption{\label{f:Bpol.r:lens} We show the theoretical B-polarization signal from lensing of E-modes (solid) and from the tensor modes for $r=0.1$ (dashed),  $r=10^{-2}$ (dotted) and $r=10^{-3}$ (dot-dashed).}
\end{figure}

As is shown in Fig.~\ref{f:Bpol.r:lens}, for  $r=10^{-2}$ the tensor B-modes are barely discernible for $\ell\lesssim 10$ while for  $r=10^{-3}$ they are nearly entirely `buried' in the lensed E-modes; only the lowest modes,   $\ell\lesssim 5$, which have large errors due to cosmic variance are higher than the signal from lensed E-modes. Nevertheless, since the lensing spectrum can, in principle, be calculated and subtracted and since it is non-Gaussian, there is not only hope but concrete 
plans~\cite{Andre:2013afa,Hazumi:2014} that future  experiments might extract B-modes down to $r=10^{-3}$. To compare  Fig.~\ref{f:Bpol.r:lens} with the BICEP2 data shown in \ref{f:Bicep2} which is given in $(\mu$K$)^2$ we have to multiply the vertical axis with $T_0^2 = (2.725\times 10^6\mu$K$)^2$. 

The discovery of B-polarization may well lead to the third Nobel Prize for the CMB.

\subsection{The CMB spectrum}
As I have mentioned in Section~\ref{s:spec} the best information we have about the CMB spectrum comes from the COBE satellite which took data in 1990, hence from an experiment which is 25 years old. Clearly, present technology could do much better. Considering the limits on spectral distortions given in Eq.~(\ref{e:spec-dist}) published by the team which has analyzed the COBE data~\cite{Fixsen:1996nj}, one may ask whether an improvement is really necessary. The answer is yes for several reasons, let me just mention the two major ones:

First, we know that the hot electrons in the reionized intergalactic medium should lead to a global $y$-distortion of the CMB of about $y\simeq 10^{-7} - 10^{-6}$.  Furthermore, the diffuse intergalactic medium is expected to generate~\cite{2000PhRvD..61l3001R}  about $y\simeq 10^{-6}$. An experiment with a sensitivity better than this would see evidence from reionisation.
As mentioned before, the $y$-distortion from individual clusters, i.e. their Sunyaev-Zel'dovich (SZ) effect~\cite{1969Natur.223..721S}, has been exploited e.g. to detect clusters~\cite{Ade:2013skr}.  See also~\cite{Bleem:2014lea} for a recent compilation of 677 SZ-selected clusters.

Furthermore, an injection of photons into the Universe happening after $z\simeq 2\times 10^6$ is no longer thermalized to lead to a blackbody spectrum, but manifests itself as a chemical potential since at these redshifts processes which change the photon number (double Compton scattering and Bremsstrahlung) are no longer active.

In addition to that, the Silk damping of acoustic oscillations in the CMB on small scales also leads to an energy injection generating a $\mu$-distortion of the order of~\cite{2012MNRAS.419.1294C}
\be\label{e:acc-mu}
\mu \simeq 1.4\frac{\De\rho_\ga}{\rho} \simeq  0.74\times 10^{-8}\,. 
\ee
For the first $\simeq$ sign we used that at redshifts $10^6>z>1100$ when Silk damping mainly occurs, photon number changing processes are no longer active so that $\De n_\ga=0$, see~\cite{2008cmb.book}.
The result~(\ref{e:acc-mu}) depends on the spectral index of primordial fluctuations on very small scales which are otherwise inaccessible to us exactly since they are damped. This represents a new way to access the primordial fluctuation spectrum from inflation on very small scales.

The details of the spectral modifications are somewhat more complicated than a simple chemical potential: at low frequencies double Compton and Bremsstrahlung
are active longer than on high frequencies and so a somewhat `frequency dependent chemical potential', $\mu(\nu)$ develops. The details of this and other heating and cooling processes of the CMB are studied in~\cite{2012MNRAS.419.1294C}.

Recently, a satellite experiment to measure the CMB spectrum at 400 frequencies from 40GHz to 6 THz named PIXIE has been proposed~\cite{Kogut:2011xw}.
Such an experiment could detect values of
\be
y \simeq 10^{-8}\,,  \quad \mu \simeq 5\times 10^{-8}  \quad \mbox{ at } 5\si\,.
\ee

An experiment of this kind would not only detect signatures from reionisation, but it would also open a new window to the primordial fluctuation spectrum on very small scales. 

\subsection{The precision of present and future CMB Boltzmann codes}
Present CMB codes announce that they are 0.1\% accurate in the relevant range of cosmological parameters. This is an amazing progress for cosmology as it allows us, with sufficiently good data, to determine cosmological parameters beyond percent accuracy. It is not so much that cosmologists want to know, e.g., $\Om_\La$ to 1\% or better, but we want to test the consistency of the standard $\La$CDM cosmological model to as good a precision as possible. 

First of all, that is what we physicists do. We test our theories to their limits. Small deviations which are only visible when measurements are sufficiently accurate can indicate flaws in the theory. For example the measured perihelion advance of Mercury is 574 arc seconds/Julian century. The theoretically calculated one within Newtonian gravity due to perturbations by the other planets is 531 arc seconds/Julian century. These calculations (all done by hand!) were very accurate and physicists knew  already around 1900 that this discrepancy of 8\% posed a real problem. This was the first indication that the Newtonian theory of gravity is not the full story.  The missing 43 arc seconds per century are due to relativistic effects and Einstein was "einige Tage fassungslos for freudiger Erregung"~(A. Einstein, letter to P. Ehrenfest, January 17, 1916) when he had done the relativistic calculation and obtained the missing 43 arc seconds~\cite{Einstein}.
Of course, even though Einstein was aware of this discrepancy, it was not what motivated him to formulate the theory of General Relativity. Nevertheless, today this is one of the crucial classical tests of General Relativity, see Ref.~\cite{Will:2005va}.

Therefore, we need very precise codes in order to be sure that a possible discrepancy is not due to inaccuracies of our calculations.. There are some doubts that the accuracy of the presently available Boltzmann codes to calculate CMB anisotropies is as good as announced. Especially, it has been shown recently that second order lensing, which is not included in these codes, can lead to changes up to 1\% in the area distance to the CMB~\cite{Clarkson:2014pda}.
This claim is especially important as a change in the area distance, $d_A\ra d_A(1+\De_d)$ implies a
 change in $h$, $h\ra h(1+\De_h)$ given by 
\be
  \De_h = \frac{d_A}{h\dd d_A/\dd h}\De_d  \simeq  -5\De_d 
\ee
for Planck values of the cosmological parameters. Hence if the background area distance is 1\% smaller than the measured one, this implies that the Hubble parameter is 5\% larger than the one inferred, assuming that the value of $d_A$ measured in the CMB is purely due to the background cosmology.

Of course, a CMB code never directly uses the distance to the CMB but its results depend on it. Therefore, if
 second order lensing can lead to 1\% effects it might also be relevant for CMB anisotropies and especially polarisation. This means that we have to modify the present Boltzmann codes to include it, see~\cite{Hagstotz:2014qea} for an attempt in this direction, see also~\cite{Bonvin:2015uha} where it is shown that the relevant effects up to second order are included in present CMB codes. Nevertheless, we have to carefully investigate whether any effect in the CMB anisotropies and polarisation might be larger than 0.1\%. In order to push precision cosmology to the next level, we have to thoroughly rethink our present Boltzmann codes.

\subsection{Large scale structure}
CMB cosmology has been tremendously successful. The reason for this is twofold. On the one hand, we have excellent high precision measurements of CMB anisotropies and polarization. On the other hand, the theoretical predictions are relatively straight forward to calculate (with the caveat mentioned in the previous section), since they are small and linear perturbation theory is quite accurate.

Can a similar program be repeated with the cosmological large scale structure (LSS), i.e. the distribution of galaxies forming clusters, filaments and voids?
At first one might be rather pessimistic: first of all, density perturbations grow large and cannot be described by linear perturbation theory. Secondly, we only see galaxies and it is not well understood how this discrete set of points traces the density field, this is the biasing problem.

Nevertheless, on large enough scales or at early times density fluctuations are small. And on large scales bias is probably linear or can be described with a few nuisance parameters. 
The gain from a precise analysis of LSS data comes mainly from the fact that, contrary to the CMB, this is a three dimensional data set. Therefore, the number of modes between a minimal, $\la_{\min}$, and a maximal wavelength, $\la_{\max}$, scales like $(\la_{\max}/\la_{\min})^3$, not like $(\la_{\max}/\la_{\min})^2$ as for the CMB. Hence even if we only
have 3 orders of magnitude in wavelength this contains in principle $10^9$ independent modes which we can add to the information from the CMB. 

It is not only, but also for this reason that there are presently several LSS surveys under way and in planning, like BOSS~\cite{Delubac:2014aqe}, DES~\cite{DES} and especially Euclid~\cite{euclid}.
In addition to the density fluctuations, the galaxy distribution which is observed in angular and redshift space contains information about the velocity field (redshift space distortions) and about the lensing potential via its deflection of the light from galaxies and other relativistic effects, see~\cite{Yoo:2010ni,Bonvin:2011bg,Challinor:2011bk}. All terms apart from the density field are  not affected by biasing and therefore may give better tracers of the matter distribution. 

Apart from the galaxy distribution, future surveys, especially Euclid, will also measure galaxy shapes which are sensitive to the shear which also determines the lensing power spectrum.
On the theoretical side, we expect significant further progress in the calculation of nonlinear aspects of clustering via $N-$body simulations, including baryon physics on small scales~\cite{Vogelsberger:2014dza} and relativistic effects on large scales~\cite{Adamek:2014xba}, or via higher order perturbation theory~\cite{Bernardeau:2001qr} and effective field theory techniques~\cite{Porto:2013qua}.

Clearly, apart from the CMB, future observations of LSS hold a lot of potential not only for precision cosmology but also for testing the theory of General Relativity in the decade to come and probably longer. The tests of General Relativity are especially important as they are on much larger scales than tests in the solar system or in binary pulsar systems.

\section{Conclusions}\label{s:con}
In this contribution I have recounted the most amazing success story of cosmology, the discovery and the analysis of the Cosmic Microwave Background. We have seen that this data not only provides us with a 'photograph' of the Universe at the very early time of about $3\times 10^5$ years after the hot Big Bang, but it contains information about the earliest stages of the Universe, probably some form of  inflation, which may have happened at an energy scale of up to  $10^{16}$GeV, before the Universe reheated and the hot 'Big Bang' happened.  The traces which such a phase of inflation has left in the CMB may even open up a window to quantum gravity, to string theory or to the multiverse.

The discovery of the CMB convinced most physicists of the hot Big Bang model: our Universe has emerged from a much hotter and denser state by adiabatic expansion and cooling. During this process small initial fluctuations have grown under gravitational instability to form the observed large scale structure. The observation of coherent acoustic peaks in the CMB fluctuation spectrum has convinced us that the  initial fluctuations actually emerged from quantum fluctuations during a phase of very rapid expansion, inflation. In other words the fluctuations in the CMB, the largest structures in our Universe, come from quantum fluctuations which have expanded and then have frozen in as classical fluctuations of the spacetime metric.

The Universe acts as a giant magnifying glass. It enlarges tiny quantum fluctuations from a very high energy phase into the largest observable structures.

While this text was finalised, the new 2015 Planck data came out, see especially~\cite{Planck:2015xua}. However, since these data are still preliminary, and since they mainly differ from the 2013 release by somewhat smaller error bars, I have not included them in this review.

\ack  I thank Martin Kunz and Malcolm MacCallum for useful discussions and Francesco Montanari for help with a figure. This work is supported by the Swiss National Science Foundation.

%\section*{References}

\bibliographystyle{iopart-num}
\bibliography{cmb}

\end{document}